\documentclass[11pt,letterpaper]{article}
\usepackage{arxiv}
\usepackage{graphicx}
\usepackage{subfig}
\usepackage{mathptmx}
\usepackage{rotate}
\usepackage{wrapfig}
\usepackage{caption}
\usepackage[T1]{fontenc}
\usepackage[english]{babel}
\usepackage{times}
\usepackage{amsmath,amstext,amsbsy,amsopn,amsthm,amsfonts,amssymb}
\usepackage{mathtools}
\usepackage{url}
\usepackage{tabularx,booktabs}
\newcolumntype{C}{>{\centering\arraybackslash}X}
\usepackage{booktabs,array,colortbl}
\usepackage{multirow}
\usepackage{algorithm}
\usepackage{balance}
\usepackage{placeins}
\usepackage{array}
\usepackage{blindtext}
\usepackage{caption}
\usepackage{capt-of}
\usepackage{booktabs}
\usepackage[small]{titlesec}
\usepackage{paralist}
\usepackage{verbatim} 
\usepackage[usenames,dvipsnames,svgnames]{xcolor}
\usepackage{algorithmic}
\usepackage{wrapfig}

\providecommand{\keywords}[1]{\textbf{\textit{Keywords---}} #1}
\begin{document}

\title{Eternal-Thing 2.0: Analog-Trojan Resilient  Ripple-Less Solar Energy Harvesting System for Sustainable IoT in Smart Cities and Smart Villages}

\author{
\begin{tabular}{ccc}
Saswat K. Ram & Sauvagya R. Sahoo & Banee B. Das\\
Dept. of	ECE  & Dept. of ECE & Dept. of CSE \\
	NIT, Rourkela, India. &  NIT, Rourkela, India. &  NIT, Rourkela, India.  \\
	saswatram01@gmail.com & sauvagya.nitrkl@gmail.com & banee.bandana@gmail.com
\end{tabular}\\\\
\begin{tabular}{cc}
\textbf{Kamalakanta Mahapatra} & \textbf{Saraju P. Mohanty} \\
\textbf{Dept. of ECE}  & \textbf{Dept. of CSE}\\
\textbf{National Institute of Technology, Rourkela, India.} & \textbf{University of North Texas, USA.} \\
\textbf{kkm@nitrkl.ac.in} & \textbf{smohanty@ieee.org}
\end{tabular}\\
}

\maketitle

\begin{abstract}
Recently, harvesting natural energy is gaining more attention than other conventional approaches for sustainable Internet-of-Things (IoT). System on chip (SoC) power requirement for the IoT and generating higher voltages on-chip is a massive challenge for on-chip peripherals and systems. Many sensors are employed in smart cities and smart villages in decision-making, whose power requirement is an issue, and it must be uninterrupted. Previously, we presented Security-by-Design (SbD) principle to bring energy dissipation and cybersecurity together through our ''Eternal-Thing''.  In this paper, an on-chip reliable energy harvesting system (EHS) is designed for IoT end node devices which is called ``Eternal-Thing 2.0''. The management section monitors the process load and also the recharging of the battery/super-capacitor. An efficient maximum power point tracking (MPPT) algorithm is used to avoid quiescent power consumption. The reliability of the proposed EHS is improved by using an aging tolerant ring oscillator. The proposed EHS is intended and simulated in CMOS 90nm technology. The output voltage is within the vary of 3-3.55V with an input of 1-1.5V. The EHS consumes 22$\mu$W of power, that satisfies the ultra-low-power necessities of IoT sensible nodes.
\end{abstract}



\section{Introduction}
\label{sec:Introduction}

Advancement in fabrication technologies and scaling down MOSFETs leads to tiny devices for various applications, including the Internet of Things (IoT) and consumer electronics. The IoT inspires every aspect of life by making things smarter with the increasing number of users. Smart cities and smart villages are now gaining more attention due to IoT for making life smarter and more advanced. The IoT consists of end node devices (sensor nodes), gateways, and cloud \cite{omairi2017power,ram2020energy}.
For an IoT node to become self-sustainable, the power hungry sensors' energy requirement should be taken care of, as the batteries are having a limited lifetime.
A smart node in IoT has the capability of sensing, processing, and communicating required information.
Designing a state of art energy harvesting system (EHS) is a demanding exercise, and factors affecting its performances are also equally important and are discussed in this paper.

%


An uninterrupted power supply from natural sources can fulfill the power requirement of an IoT smart node, which should cause less damage to society and the environment.
The smart node has to monitor activities without any interruption for efficiently gathering information. An uninterrupted supply for the IoT node is a must regardless of deployment
\cite{shao2009design}. Traditionally, batteries are used as a supply, but it has specific adverse effects on humanity from the disposal point and the substance side. Rechargeable batteries/supercapacitors were investigated and are found to be economical and safe \cite{ram2020energy}. Harnessing natural energy and storing in supercapacitors is a suitable alternative.
Towards energy savage, clean energy, and environment-friendly approaches, solar energy is preferred. The enough presence, and absence of mechanical parts in its conversion to electrical is an advantage among various available natural resources as solar \cite{shao2009design,ram2019sehs}, thermoelectric \cite{omairi2017power,carreon2014boost}, radiofrequency (RF), piezoelectric, and wind \cite{omairi2017power}. Solar power is affected by low conversion efficiency but can be overcome by using suitable conditioning techniques. The charge pump is preferred as converter for boosting solar output, and are suitable for on-chip implementation \cite{shih2011inductorless} to drive loads for various applications \cite{das2019spatio,ram2019ultra},
\cite{kim2011regulated}. The PV cell performance is depending on various parameters like illumination intensity, temperature levels. An IoT smart node contains of energy supply (solar), harvest system, SoC, sensors, and transceivers  \cite{liu2015highly}. There are possibilities of attacks due to various adverse conditions in the energy harvesting module to degrade its performance, and it should be investigated.

The remaining of the paper is organized within the following manner: Section \ref{sec:Smartcityandvillage} presents our idea of eternal thing for improving the sustainability in smart villages and smart cities. Section \ref{sec:ResilientIoT} discusses the reliability issues and resilient IoT node as Eternal-Thing. Section \ref{sec:Relatedwork} describes the related prior research. Section \ref{sec:Novel} presents the novel contribution of this research paper. Section \ref{sec:EHS} elaborates the PV harvesting system. Section \ref{sec:AnalogTrojandetection} discusses the reliability issues with detection and mitigation mechanisms. Section \ref{sec:Experimentalresults} presents the experimental results, and at last, section \ref{sec:Conclusion} concludes the paper.

\section{Security-by-Design Sustainability in IoT for Smart Cities and Smart Villages}
\label{sec:Smartcityandvillage}

\begin{figure}[htbp]
	\centering
	\includegraphics[width=0.98\textwidth]{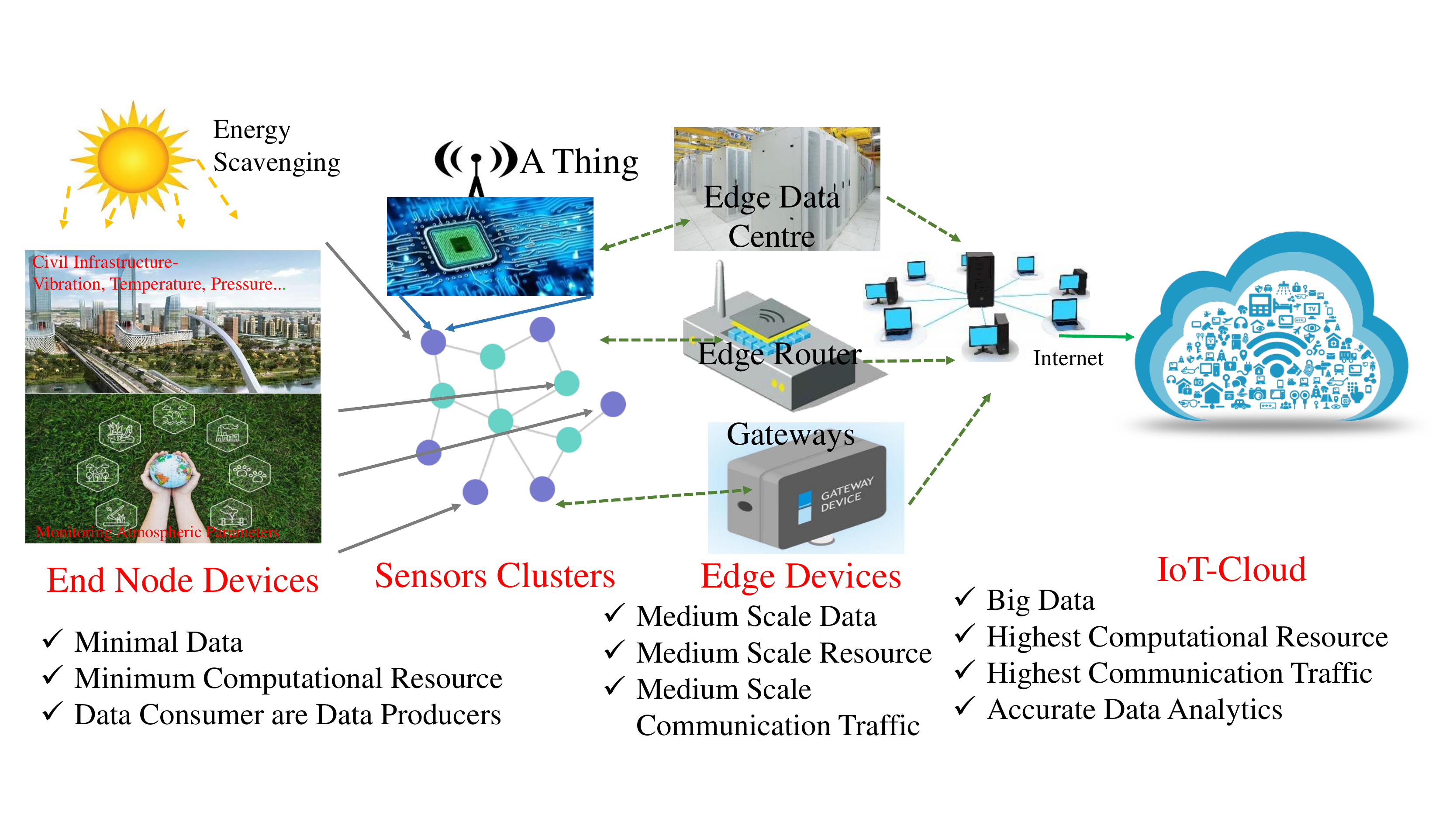}
\caption{IoT Computing - A requirement for Smart Villages and Smart Cities \cite{ram2020energy}}
	\label{fig:IoT_Paradigm}
\end{figure}

The continuous monitoring at end node makes the devices energy hungry \cite{8287053}.
The budget is increased by use of conventional batteries as it needs to be replaced regularly, and the lifespan of them is limited.
The continuous replacement of battery increases the budget as it get drained out soon. 
The secure and sustainable energy harvesting technologies for IoT in 
smart villages and smart cities are in demand as depicted in Fig. \ref{fig:IoT_Paradigm}.

\subsection{Smart Cities}

By 2050, 70\% of the world population migrate to cities due to the technical advancement in smart cities, as worldwide smart cities are making substantial progress \cite{7539244}.
Smart city engages the key industries with applications like good governance, good utility, improved quality, smart buildings, and good surroundings \cite{zanella2014internet}. The performance of assorted infrastructures are often increased by use of IoT and AI-driven information analytics smart cities. Challenges in smart cities includes the design and operation cost, power requirement of devices, reliability and security, big data analytic and disaster resilience \cite{gubbi2013internet, rostirolla2017elcity, Alladi_MCE_2020-Mar}.

\subsection{Smart Villages}

The use of renewable energy with current technical developments can influence an estimated 940 million people worldwide. The smart village and smart cities are having some common things but there are some distinct functionalities are also present \cite{ram2020energy, 9333580}. Efficient use of technologies as information and communication technology (ICT), geography information system (GIS), IoT-cloud, IoT-edge, and remote sensing  may be employed in building a smart village \cite{davies2020iot, 9153927}.

The challenges for sustainable IoT are the following \cite{8287053}: (a) High Energy Requirement (b) Security, Reliability, and Privacy, (c) Energy and Cost, and (d) Latency and Bandwidth. The smart services include the following \cite{ram2020energy}: (a) Smart Energy, (b) Waste Management and Air Quality, (c) Smart Transportation, (d) Smart Healthcare, (e) Smart Home, (f) Noise Monitoring and Traffic Congestion, and (g) Structural Health of Buildings. These challenges and services involve many sensors, and providing uninterrupted supply to these devices is a challenge.

\subsection{Security-by-Design (SbD) or Secure-by-Design (SbD)}

Security-by-Design (SbD) is a design paradigm that forces cybersecurity as a design axes in the early design space exploration of circuits and systems \cite{8977808}. SbD ensures cybersecurity doesn't need retrofitting at the later stage which may be expensive. SbD also ensures that cybersecurity doesn't put energy or battery overheads on the hardware and system. One of the earliest hardware that demonstrated such SbD paradigm is our Secure digital Camera (SDC) which has builtin cybersecurity components, but has no energy overheads \cite{1512191, MOHANTY2009468, 8263198}. We have presented World's first low-power watermarking chip that provides data protection while consuming a very battery or energy resource \cite{1632351}. We further deployed SbD in Internet-of-Medical-Things (IoMT) in our physical unclonable function (PUF) based lightweight cybersecurity of medical devices with power consumption of few $\mu$Watt \cite{8752409}.

In our ongoing research, we are mainly focusing on security-by-design (SbD) or secure-by-design (SbD) concepts for electronic systems \cite{ram2020energy}. We explore the security and reliability in ICs for performance improvement of energy harvesting systems that supply the end node devices used in IoT. The need for energy harvesting in different application domains targeting the smart village and smart cities are possible \cite{ram2020energy}. In \cite{ram2020energy} the factors that influence the living in smart villages and smart cities, along with the technical aspects, were well presented. In \cite{ram2020energy}, the energy perspective for the smart village and smart cities with security and reliability are explored. The reliability of ICs is an important aspect and is discussed in this paper for the sustainability of energy harvesting systems.
In \cite{ram2020eternal} we used physically unclonable functionality (PUFs) as security primitive to secure our EHS IC and presented the concept of ``Eternal-Thing''. The IC also contains an aging sensor, which is capable of detecting the recycled ICs.

\section{The Issue of Reliability and Our Vision of Resilient IoT Node as Eternal-Thing 2.0}
\label{sec:ResilientIoT}

The category that most influences the hardware industry is the recycled integrated circuits (ICs). It is estimated that 80\% of the forged/counterfeited ICs sold worldwide are recycled ones. The use and nature of short-lifetime gadgets lead to an increase in e-waste and recycled products. These wastes are again recycled as new ICs by the adversaries, which mainly influences two aspects, i.e., security and reliability. In our previous work \cite{ram2020eternal}, the main focus is on the security of the EHS IC and detecting recycled EHS IC.
In this work, we are mainly concentrating on the harvesting system design with reliability and security issues along with its mitigation.  We can not ignore these as the IoT node has to be ON for more extended periods and are challenged at each stage, beginning from the fabrication till deployment by various adverse scenario created by adversaries. The different reliability issues in the EHS are the following: (a) Presence of ripples at the output, (b) Intentional aging, and (c) Adding suspicious circuits in the existing circuitry.


\subsection{Ripples at output}

The presence of ripples at the output of the EHS is equally hazardous. It may cause heating of devices, noise, distortions and can reduce the lifespan of these devices. These ripples should be minimized for designing an efficient and reliable energy harvesting system. 
These ripples are due to the following reasons: (a) optimizing the capacitor size and the output current by fast switching of MOS (b) low on-resistance (c) rapid charging and discharging of pump capacitors.
Alternative solutions should be investigated to address this problem, and is well explained in Section \ref{sec:AnalogTrojandetection}.
By proper investigation it is found that the ripples at output of DC-DC converter depends upon the oscillation frequency ($f_{ossc}$). 
Any variation in the $f_{ossc}$ increases the ripple at the output.

\subsection{Resilient against Attacks}

Attacker uses different Trojans to affect the reliability of the IC. To affect the reliability of EHS, different types of attacks can be made by adversary.
\begin{description}
\item[$\bullet$ ] The oscillating frequency ($f_{osc}$) of ring oscillator (RO) is one of the key parameters, which predicts the reliability, the attacker may try to age the RO of EHS by subjecting the RO section to very high temperature.
\item[$\bullet$ ] The Analog Trojans like A2, uses the inherent characteristics of the circuits like charge sharing and capacitative coupling to disturb the circuit operation \cite{guo2019capacitors,yang2016a2}. 
\end{description}
  

\subsection{Analog Trojan (A2) and its influence}

The boost in the IC market and its globalization put threat of hardware Trojans (HTs) among various available security issues in integrated circuits. 
Current HTs detection techniques assume HTs are only consists of digital circuits. 
In recent days to affect the reliability of circuits, Analog Trojans are used. 

The empty spaces in the layout can be used by this type of Trojans to affect the performance and behavior of the circuits. The detail analysis of charge enabled attacks and its detection is discussed in subsequent sections.

\subsection{The Proposed Eternal-Thing 2.0}

\begin{figure}[htbp]
	\centering\includegraphics[width=0.90\textwidth]{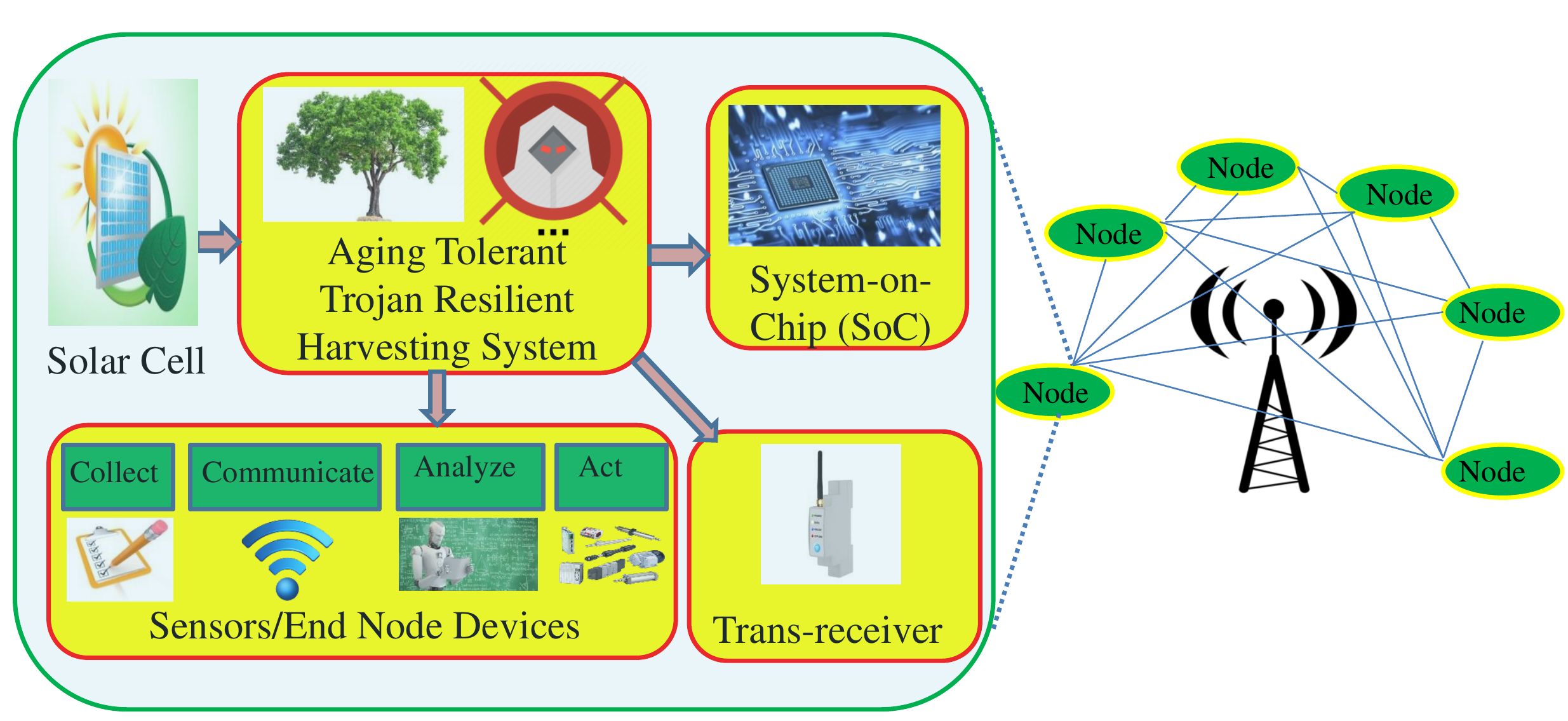}
\caption{An Reliable Trojan Resilient IoT Smart node - The ``Eternal-Thing 2.0''.}
	\label{fig:Smartnode}
\end{figure}

We envision the end node of the internet of things (IoT) with energy harvesting and security or reliability capability as ``Eternal-Thing 2.0''. Fig. \ref{fig:Smartnode} shows the high-level structure of a sustain reliable Trojan resilient IoT smart node.
As depicted in Fig. \ref{fig:Smartnode}, the supply to end node devices, SoCs and trans-receivers are supplied from the EHS. Any deviation in supply cause malfunction in these devices, so the harvesting system should be resilient and reliable enough to a different type of attack caused by an adversary. The harvesting system's design addressing these unavoidable things motivates us to design a reliable Trojan resilient harvesting system (``Eternal Thing 2.0''). As per the author's knowledge, the security, and reliability of EHS were never discussed in earlier literature.

\section{Related Prior Research}
\label{sec:Relatedwork}

The renewable sources are now preferred as an input source but are limited by conversion efficiency.
There are various control techniques for different type of input energy sources are available in literature for energy harvesting system \cite{omairi2017power, shao2009design, carreon2014boost, kim2013energy, mondal2017chip, ram2018energy, ram2019ultra, ram2020solar}. A multi source energy harvesting system was discussed in \cite{estrada2019fully, abuellil2019multiple, ma2019sensing}, where an algorithm is used for detecting the input with maximum power. Authors in \cite{chew2018power} presents a piezoelectric harvester, which uses DC-DC converter and is able to supply the IoT sensors. Photovoltaic (PV) is a prominent energy source for its wide local availability, sustainability, environment-friendly nature, no moving parts, and no carbon emission.
Researchers have discussed various voltage boosting techniques in the literature \cite{chang2018modeling}.
By adopting suitable conditioning techniques, energy can be extracted for supplying the computational load and is stored for blackout periods in rechargeable battery/supercapacitors \cite{liu2015highly}. For high-voltage and high-current drive supercapacitors are the preferred. It covers a broad vary of applications as IoT, shopper product, white merchandise, workplace automation, long run battery backup and energy harvesting and are well presented in \cite{sengupta2018supercapacitors}. The self-sustainable energy harvesting system design for IoT nodes uses DC-DC converters for boosting as the output achieved is non-conditioned. The charge pumps are widely accepted for on-chip implementation and low voltage applications \cite{biziitu2017improving, ballo2019review}. Various maximum power point tracking (MPPT) algorithms are used in literature \cite{omairi2017power, qian2017sidido, ram2020eternal} to achieve maximum power. Yue et al. in \cite{yue2019charge} presented the use of supercapacitor as energy storage for solar based harvesting system towards battery less design. 
There are various aging mechanism exists as Bias Temperature Instability (BTI), Hot Carrier Injection (HCI), Electromigration (EM), TDDB etc., causes continuous and irreversible degradation in ICs with time. These aging based parameters are used by researchers 
for the detection of reliability in ICs \cite{zhang2013design, guin2015design}. The threshold voltage ($V_ {t}$) is the most influenced parameter by aging in MOSFETS.
The threshold voltage degradation of MOSFETS will increase with aging duration. In \cite{sahoo2018novel, zhang2013design, guin2015design}, authors discussed aging tolerant methodologies for sustainable operation of circuits. The summary of the related work is conferred in Table \ref{tab:Relatedwork}.


\begin{table*}[htbp]
	\centering
	\caption{Related Works}
	\label{tab:Relatedwork}
	\setlength{\tabcolsep}{0.8em} 
		\begin{tabular}{|l|l|l|l|}
			\hline
			
			\parbox[c]{1.5cm}{\normalsize }{\raggedright \textbf{Works}} & \parbox[c]{6.0cm}{\normalsize }{\raggedright \textbf{Methodology Used}} & \parbox[c]{7.0cm}{\normalsize }{\raggedright \textbf{Salient Features}}    
			\\ 	
			\hline
			\hline
			
			\parbox[c]{1.5cm}{Shao, et al. \cite{shao2009design}}&\parbox[c]{5.0cm}{Variable switching frequency to control charge pump (CP)} &\parbox[c]{7.0cm}{\smallskip$\Rightarrow$ Inductor less Design.
				\\$\Rightarrow$ MPPT achievement by varying Switching Frequency.
				\\$\Rightarrow$ Extra circuitry in Variable switching frequency.
				\smallskip
			}\\
					\hline
			
			\parbox[c]{1.5cm}{Carreon, et al. \cite{carreon2014boost}}&\parbox[c]{6.0cm}{Matching impedance dynamically with Thermoelectric generators (TEG)} &\parbox[c]{7.0cm}{ \smallskip $\Rightarrow$ TEG as an alternative input.
				\\$\Rightarrow$ TEG output needs conversion to DC adding more cost.
				\smallskip
			}\\
			\hline
			
			\parbox[c]{1.5cm}{Kim, et al. \cite{kim2011regulated}}&\parbox[c]{6.0cm}{Regulated CP with Optimum Power Point Algorithm (OPPT)} &\parbox[c]{7.0cm}{ \smallskip $\Rightarrow$ Inductor less design.
				\\$\Rightarrow$ OPPT for MPPT for Indoor light conditions.
				\smallskip
			}\\
					\hline
			
				\parbox[c]{1.5cm}{Shih, et al. \cite{shih2011inductorless}}&\parbox[c]{6.0cm}{CP with BGR output Controller} &\parbox[c]{7.0cm}{\smallskip $\Rightarrow$ Four-phased clocking in charge pump design.
				\\$\Rightarrow$ Band gap reference circuit for regulation.
				\\$\Rightarrow$ Clock generation scheme is complicated.
				\smallskip
			}\\
			\hline
			
			\parbox[c]{1.5cm}{Kim, et al. \cite{kim2013energy}}&\parbox[c]{6.0cm}{Successive Approximation Register (SAR) MPPT with Active and Power down Mode} &\parbox[c]{7.0cm}{ \smallskip $\Rightarrow$ For low power consumption SAR MPPT is used in Indoor light conditions.
				\\$\Rightarrow$ Application restricted to indoor lightning.
				\smallskip
			}\\
			\hline
			
			\parbox[c]{1.5cm}{Mondal, et al. \cite{mondal2017chip}}&\parbox[c]{6.0cm}{Adaptive MPPT for harvesting System} &\parbox[c]{7.0cm}{ \smallskip $\Rightarrow$ Harvesting system design using current starved VCO (CS-VCO) for frequency tuning in MPPT.
				\\$\Rightarrow$ Additional circuits for frequency adjustment.
				\smallskip
				
			}\\
		
					\hline
			
\parbox[c]{1.5cm}{Ram, et al. \cite{ram2020eternal} (Eternal-Thing)}&\parbox[c]{6.0cm}{Secure solar energy harvesting system with recycled IC detection} &\parbox[c]{7.0cm}{ \smallskip $\Rightarrow$ Selfsustainable and secure harvesting system design.
				\\$\Rightarrow$ Low power switching scheme for converter.
				 \\$\Rightarrow$ Aging sensor used for counterfeited IC detection.
				\smallskip
			}\\
\hline
\parbox[c]{1.5cm}{\textbf{Current Paper (Eternal-Thing 2.0)}}&\parbox[c]{6.0cm}{\textbf{Analog Trojan resilient ripple-less reliable ultra low-power energy harvesting system}} &\parbox[c]{7.0cm}{ \smallskip $\Rightarrow$ Reliable solar energy harvesting system resilient to Analog Trojan.
				\\$\Rightarrow$ Converter uses charge pump as a voltage tripler by cascading doublers for better efficiency.			\\$\Rightarrow$ Detailed discussion on ripples at output and mitigation.
				\\$\Rightarrow$ Aging tolerant ring oscillator used for reducing the ripple such that the life time of the load can be increased.
				\smallskip
			}\\
			\hline
	\end{tabular}
\end{table*}

Once the EHS is designed, its reliability in term of ripple mitigation at the output and influence of Analog Trojan is equally important as security of the EHS. The security of EHS is discussed in our paper eternal thing \cite{ram2020eternal}. Further designing an aging tolerant Trojan resilient EHS is a challenge. The clock frequency degradation due to aging and type of attack due to Analog Trojan in EHS needs to be investigated and proper mitigation technique should be addressed.


In our previous work, we have addressed the security related issues for IoT smart node for sustainable IoT and called in ``Eternal-Thing'' \cite{ram2020eternal}. The physically unclonable functionality (PUF) was used as a security shield to address the security of EHS, and for detection of counterfeited IC, aging sensor was used. The PUF metrics were analyzed and were found satisfactory. The attack due to Analog Trojans (Intentional temperature variation and A2 Trojan) and presence of ripples at the output of converter in EHS insist us to investigate the cause of it and its mitigation, which is presented in the current paper along with EHS design as ``Eternal-Thing 2.0''.

\section{Novel Contributions of the Current Paper}
\label{sec:Novel}

The current paper addresses an unified energy harvesting system design that analyzes the following:
(a) ripple content at the output, 
(b) mitigation technique for countering the attack in RO section made by intentionally varying the temperature, and (c) detection of Trojan causing unwanted reset in the MPPT unit.  
The proper analysis and mitigation of attack due to Analog Trojan and ripples (causes due to aging) is experimented to design an reliable ripple-less solar harvesting system for sustainable IoT. 


\subsection{Research Question and Challenges Addressed in the Current Paper}

The design of a harvesting system itself is a challenging task. Choosing a suitable source of energy and its conditioning
needs a lot of effort. As per the deployment and usage, 
the aging effect is a concern along with security.
The reliability degradation of the EHS IC is due to several types of attacks.
These issues arise many questions, and these challenges are well addressed by the authors in this  research paper.

\subsection{Proposed Solution of the Current Paper}

The review based on various researchers' work and their outcomes encourages us to design a reliable EHS. There are multiple types of energy sources used by different researchers. The solar energy is proved to be the most economical and widely available energy source for harvesting purposes. 
The impedance matching between the PV cell and charge pump is a  concern for selecting suitable MPPT, that can be implemented on-chip. The perturb \& observe (P \& O) algorithm is appropriate with low hardware cost and ease of design for MPPT. The capacitor value modulation (CVM) scheme is adopted as the variable frequency for tuning MPPT needs additional control circuits, thereby consuming more area on-chip.

To counter the attack due to analog Trojan (intentionally varying the temperature and A2 Trojan) and the ripple analysis with proper mitigation for a ripples-less reliable EHS is designed in this paper for sustainable IoT. 
Since an aging tolerant RO is used in our design, which shows resilient against intentional aging by adversary.

\subsection{Novelty of the Proposed Solution}

The systems discussed in the earlier literature may suffer from ripples at the output and attack due to Trojan. None of the researchers addressed this issue and its mitigation. The contribution of this paper is as follows:
\begin{description}
\item[$\bullet$ ] Design of a novel ultra-low-power (ULP) self-sustainable, reliable solar energy harvesting system (PV-EHS).
\item[$\bullet$ ] A novel aging tolerant mechanism is implemented to improve the reliability of EHS.
\item[$\bullet$ ] A novel methodology is proposed to counter the effect of Trojan caused by intentionally increasing the temperature of RO and Trojan (A2) detection technique for EHS.

\end{description}


\section{Proposed Reliable Energy Harvesting System Design}
\label{sec:EHS}

\begin{figure*}[htbp]
	\centering\includegraphics[width=0.98\textwidth]{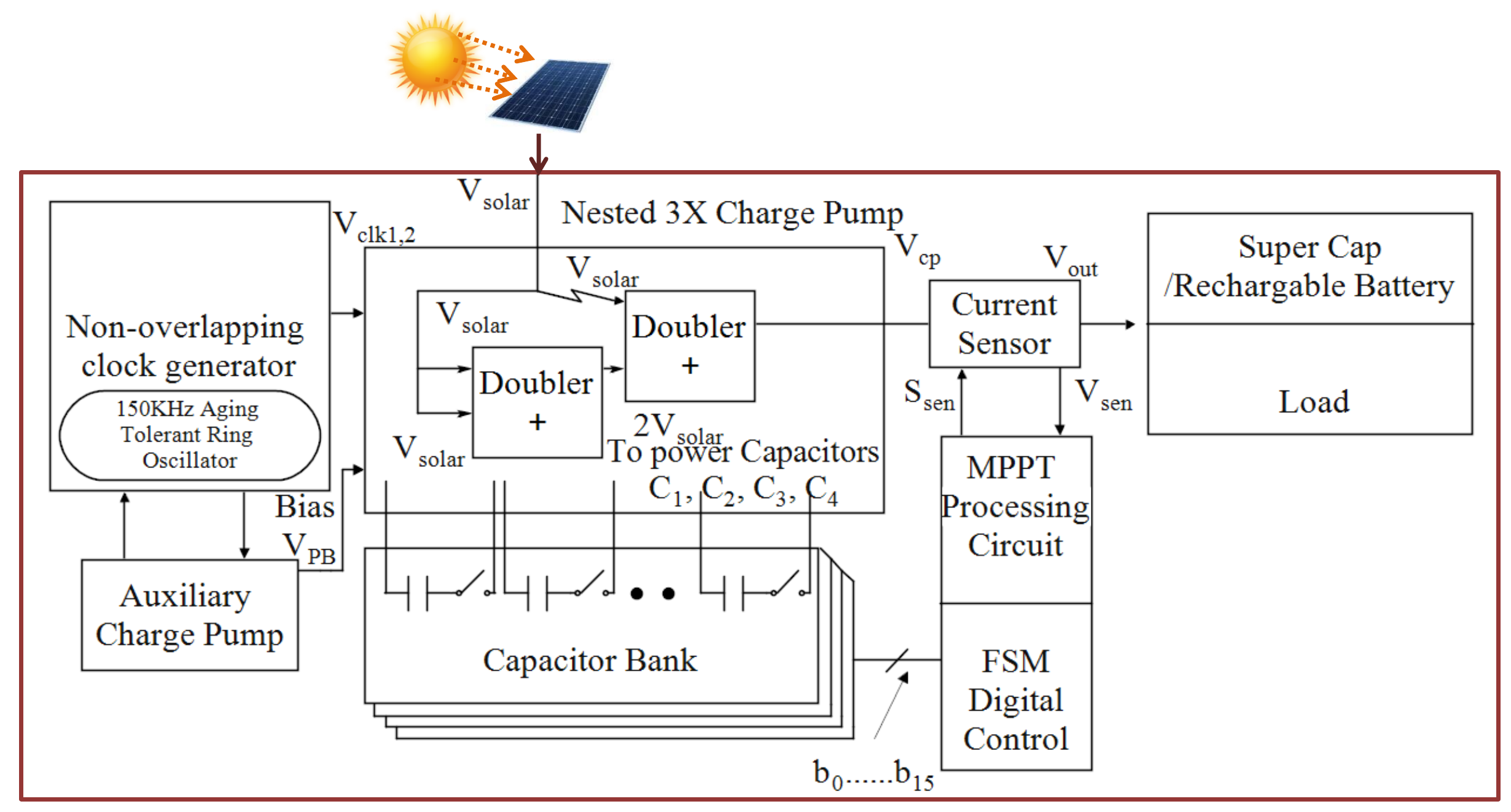}
	\caption{Block Diagram of Proposed Reliable Solar Energy Harvesting System.}
	\label{fig:EHS}
\end{figure*}

Fig. \ref{fig:EHS} depicts the proposed reliable Solar-EHS, which can convert the lower-level solar voltage to higher-level voltages using DC-DC converters to drive the computational load and rechargeable battery. The control section consists of a digital controller (FSM), which controls the current sensor and MPPT module's operation. The proposed PV-EHS is designed using (a) reliable ring oscillator, (b) a non-overlapping clock generator, (c) level shifter, (d) auxiliary charge pump, (e) current sensor, and (f) MPPT module with digital controller.
The proposed EHS is similar to the existing harvesting systems discussed in literature with further modifications to achieve improved boosting, effective MPPT module through CVM and better reliability. 
Higher reliability is achieved using aging tolerant ring oscillator and Trojan detection with mitigation techniques, which is discussed in detail in subsequent sections.

\subsection{Non-overlapping clock generator (NOCG) with level shifter (LS) and auxiliary Charge Pump (ACP)}

The reliable ring oscillator delivers the desired clock frequency (150 kHz) that is fed to a non-overlapping clock generator (NOCG).
The NOCG generates clocks for the auxiliary charge pump (ACP) and therefore the level shifter (LS) having a reduced shoot-through result. The LS, and ACP offer the switch signals and alternative biases required for voltage booster. The level shifter is supposed for reinforcing the clock amplitude, that ultimately reducing the losses because of charge transfer.

 \begin{figure}[htbp]
 	\centering
\includegraphics[width=3.7in]{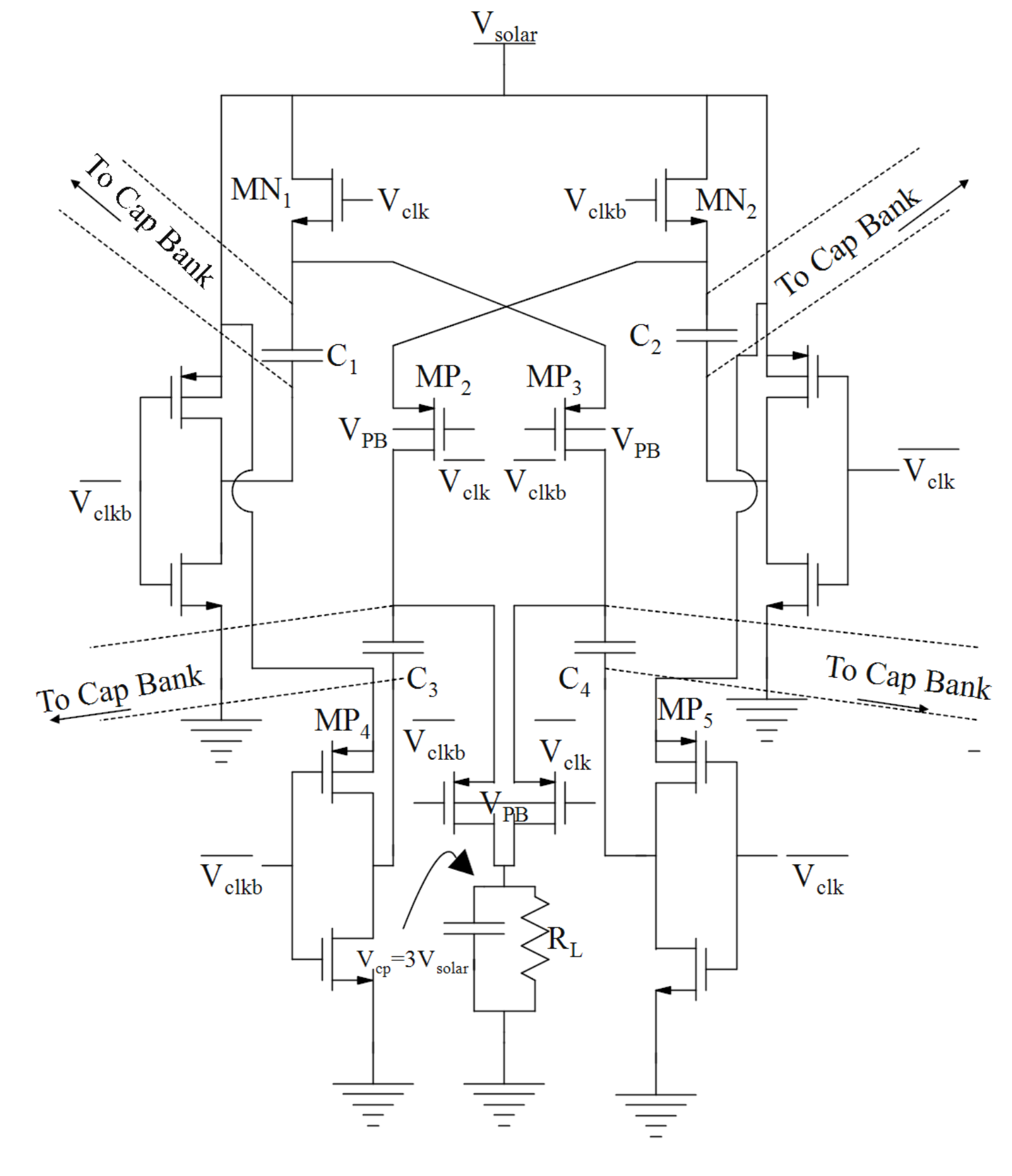}
\caption{Circuit diagram of Charge Pump as a Voltage Tripler.}
\label{fig:TRIPLER}
\end{figure}

\subsection{Charge Pump as a Voltage Booster}

Fig. \ref{fig:TRIPLER} shows a charge pump circuit diagram applied as a voltage tripler. The clocks boosts the solar voltage in the charge pump to charge the solar energy across the capacitors $C_{u}$ (which includes power capacitors $C_{1}$, $C_{2}$, $C_{3}$, $C_{4}$) and level-up the negative plate by the same potential.
The boosting employing a voltage doubler device isn't enough to get adequate provide for the sensors utilized in IoT.
The DC-DC converter used has a conversion ratio (CR) of three (CR=3). The cross-coupled MOSFETs $MN_{1}$ and $MN_{2}$ in Fig. \ref{fig:TRIPLER} are driven independently with high voltages, thereby reducing the conduction resistance.

The impedance of the voltage tripler can be expressed by the following expression \cite{liu2015highly}:
\begin{equation}
\label{eq:Impedance}
{Z_{cp}} = \frac{{{V_{solar}}}}{{{I_{in}}}} = \frac{1}{{2{f_s}{C_u}}}\frac{{1 + \alpha }}{{\left( {3 - \frac{{{V_{out}}}}{{{V_{solar}}}}} \right)\alpha }} .
\end{equation}
\par
Eqn. \ref{eq:Impedance} shows that impedance of the CP is inversely proportional to $C_{u}$. $\alpha$ is the capacitor ratio between stages of CP. The capacitor $C_{u}$ is coupled to programmable cap banks for impedance matching using CVM.
The small signal model of CP gives:
\begin{equation}
\label{eq:Smallsignal}
\left[ {2{V_{solar}} - \left( {{V_{out}} - {V_{solar}}} \right)} \right] \times \alpha {C_u} = \frac{1}{2} \times \left( \frac{{T \times {V_{out}}}}{{{R_L}}} \right).
\end{equation}
In the above expression, $T$ is the switching period and $R_{L}$ is the load of the SoC. 
By rearranging equation \ref{eq:Smallsignal}, we obtain the following expression:
\begin{equation}
\label{eq:MPP}
{V_{solar}} = \left( {\frac{1}{2}*\frac{T}{{{R_L}}}*\frac{1}{{\alpha {C_u}}} + 1} \right)*\frac{1}{3}*{V_{out}}\underrightarrow{Match}{V_{MPP}}
\end{equation}

From Eqn. \ref{eq:MPP}, it is clear that the MPPT is obtained by varying the frequency $f$ and capacitor $C_{u}$ as the frequency is constant here, so a variable $C_{u}$ is proposed. The power capacitors are digitalized as cap banks, as it consumes less power with reduced noise.
The capacitors can be directly implemented on-chip due to technological development in fabrication technologies. 
The power conversion efficiency (PCE) is given by the following expression:
\begin{equation}
\label{eq:PCE}
PCE = \frac{{{V_{out}}}}{{{V_{solar}} \times CR}} \times 100\% 
\end{equation}

\subsection{MPPT Module}

\begin{figure}[htbp]
\centering
\includegraphics[width=3.5in]{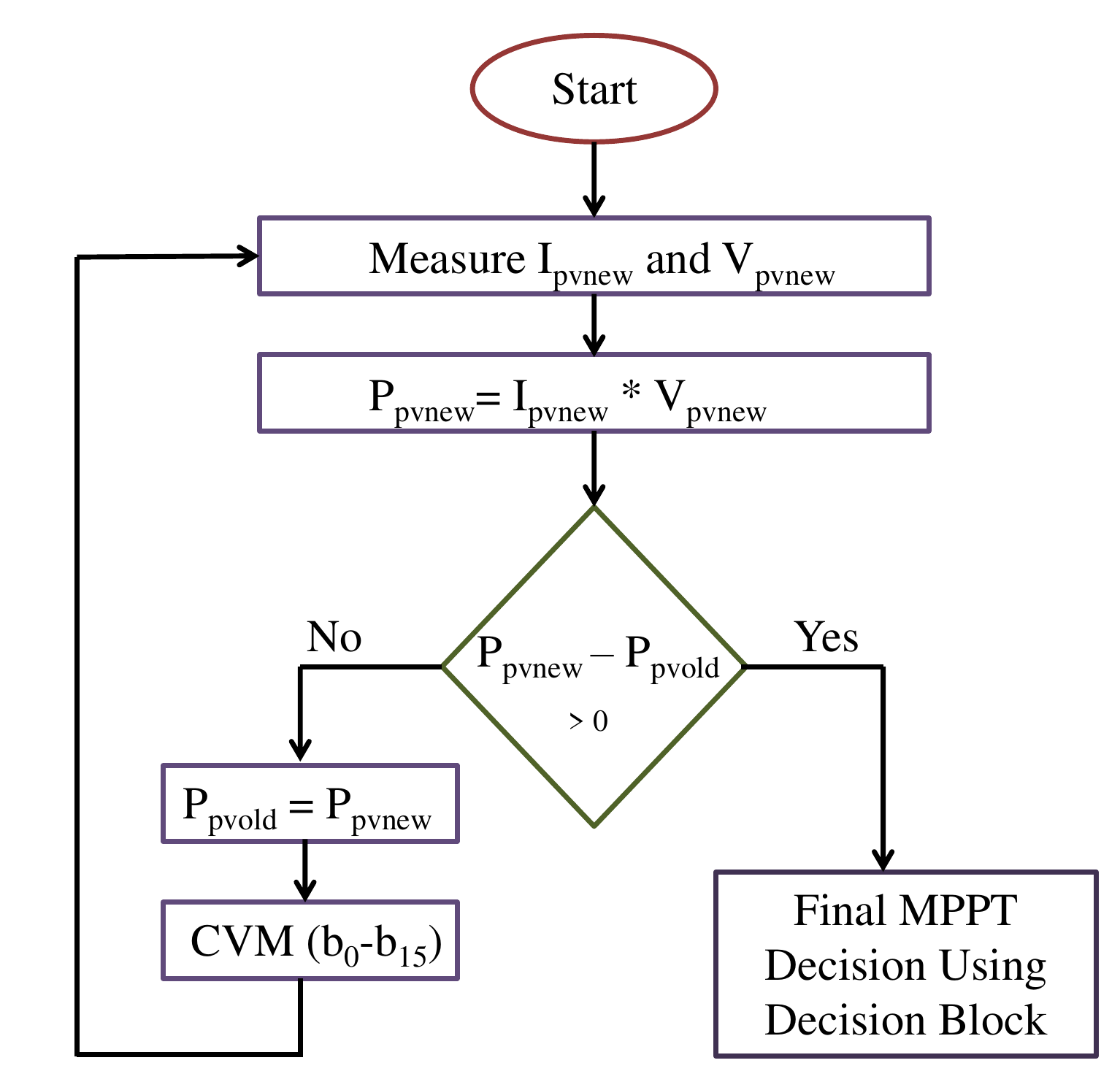}
\caption{Conventional P \& O Algorithm using CVM.}
\label{fig:Conv_MPPT}
\end{figure}

The variation in irradiance level and temperature degrades the PV-cell performance. An energy efficient Perturb \& Observe (P \& O) MPPT technique is used to extract the maximum power, which senses the voltages after capacitor value modulation (CVM) and then processes it through iterations for taking a final decision on MPPT achievement. The P \& O algorithm with CVM is depicted in Fig. \ref{fig:Conv_MPPT} in which a digital controller controls the MPPT procedure.

\begin{figure}[t]
\centering
\subfloat[Flow chart of P \& O algorithm with procedure.]{\label{fig:MPPTALGO_FSM1}\includegraphics[width=0.75\textwidth]{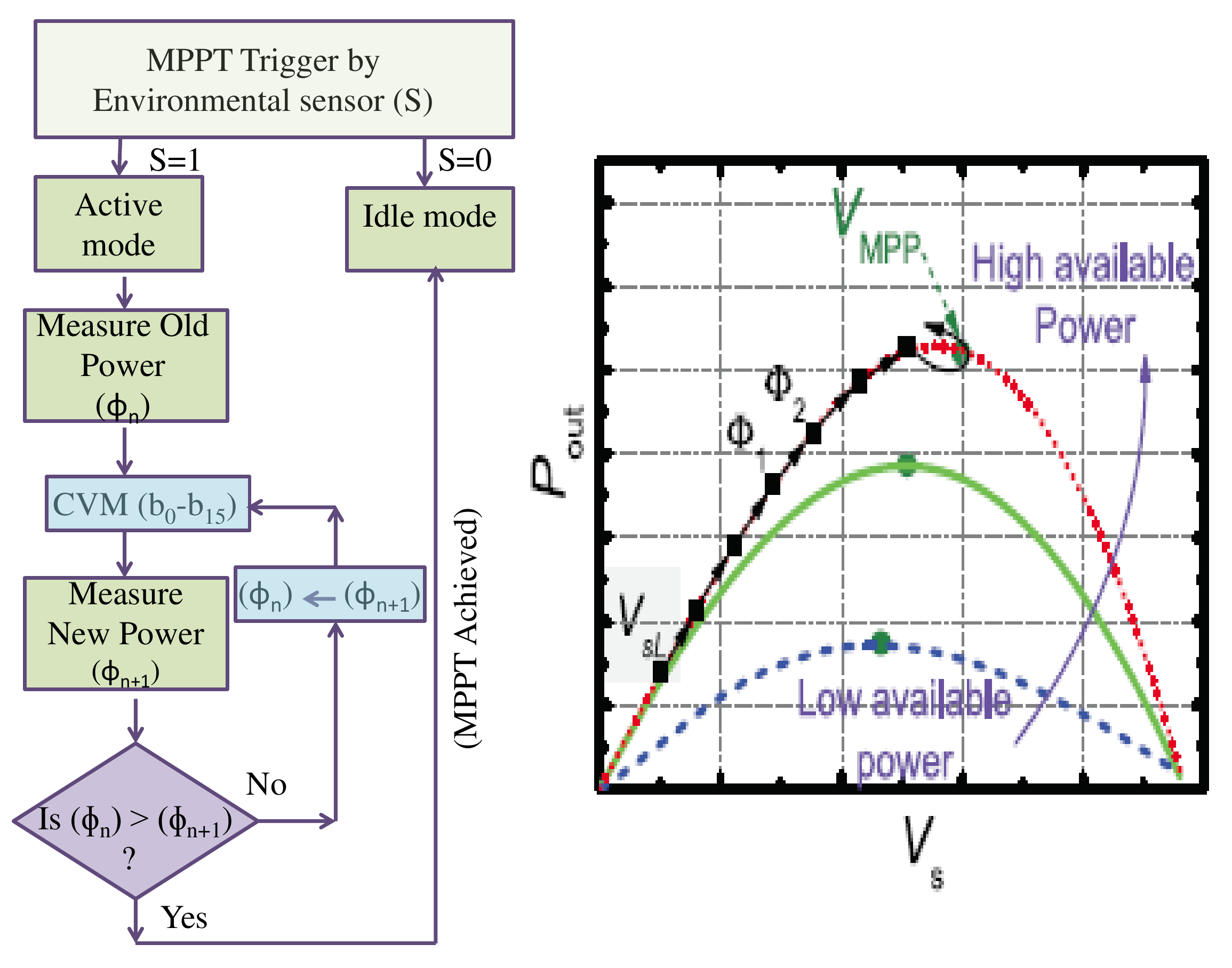}}\\
\subfloat[Finite State Machine (FSM) for MPPT.]{\label{fig:MPPTALGO_FSM}\includegraphics[width=0.70\textwidth]{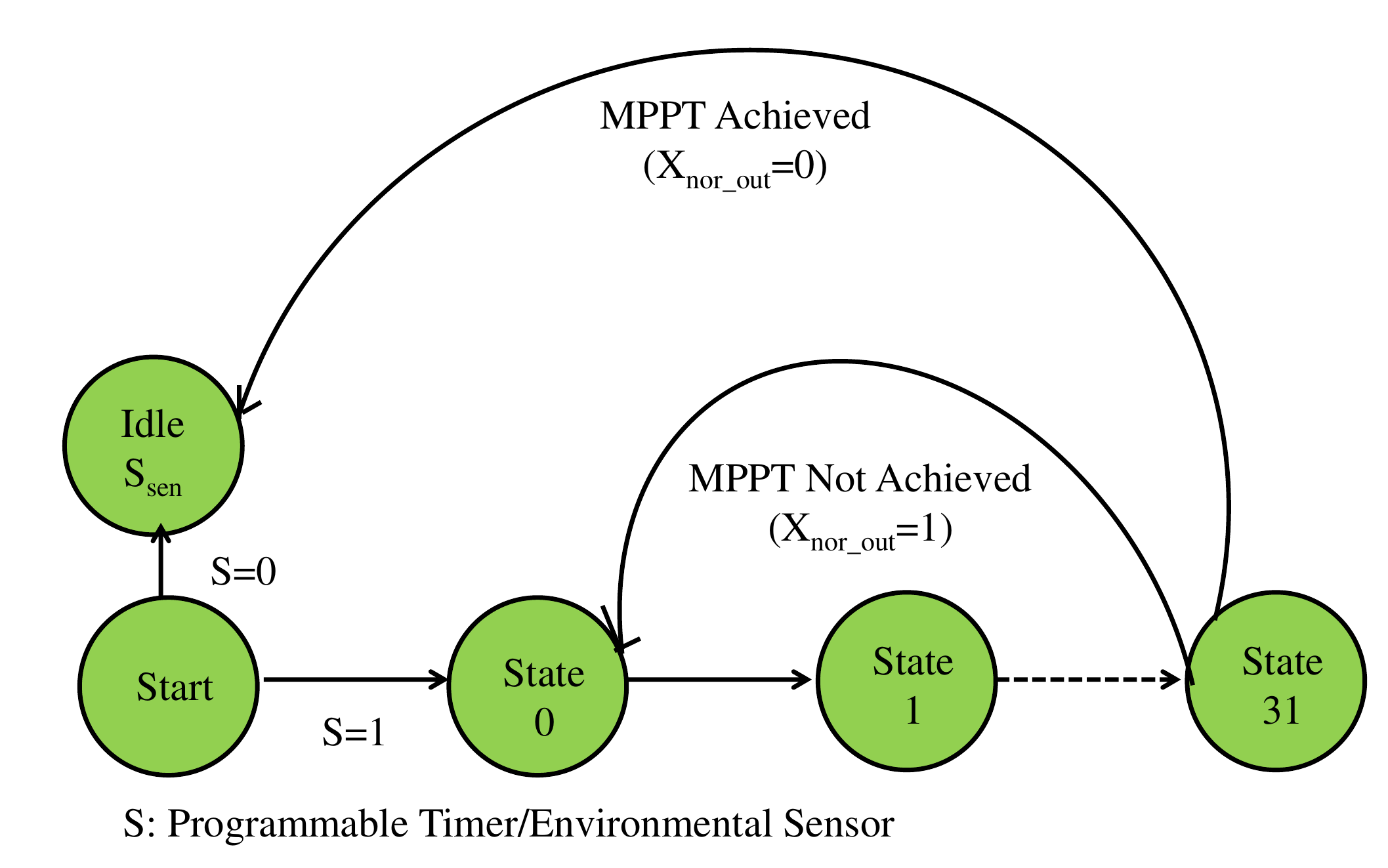}}\\
\caption{Perturb \& Observe MPPT Procedure.}
\end{figure}

\begin{figure}[t]
\centering
\includegraphics[width=0.60\textwidth]{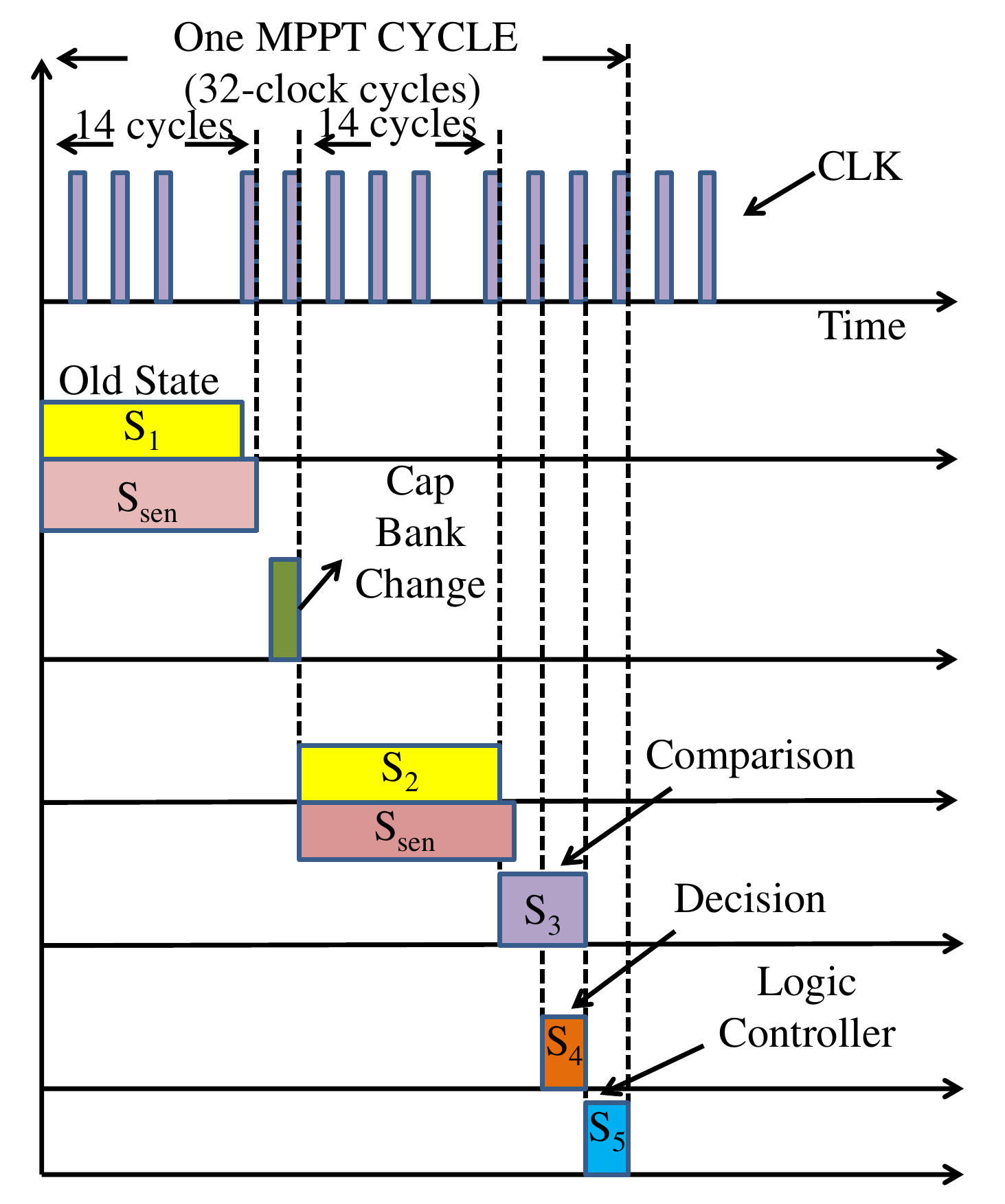}
\caption{Timing Diagram for MPPT Procedure with associated Control Signals.}
\label{fig:Timing}
\end{figure}

Fig. \ref{fig:MPPTALGO_FSM1} depicts the flowchart and procedure of the P \& O algorithm. An environment sensor triggers the MPPT procedure by enabling signal $S$. The period till $S$ is zero; the solar cell supplies load and rechargeable battery, but when the signal $S$ goes high, the MPPT circuit is triggered. The total current during the sensing phase is passed to the sample and hold circuits at equal intervals for storage, comparison, and final decision. The finite state machine (FSM) for one MPPT cycle is shown in Fig. \ref{fig:MPPTALGO_FSM}. A digital controller consisting of FSM and generates the necessary control signals for the entire MPPT process. The timing diagram in Fig. \ref{fig:Timing} explains the status of the signals generated by the controller \cite{ram2020eternal}.

A current sensing element is employed to sense the ability info from the output of the DC-DC converter.
The FSM signals control the current sensor, in this design $S_{sen}$ and $S_{senbar}$. When $S_{senbar}$ is low, the converter's total current is used for sensing, and at that instant, the supercapacitor provides the reference current to the current sensor. The sensing voltage $V_{sens}$ is obtained across $R_{SEN}$ resistor. When the MPP is achieved, $S_{sen}$ is low, and the current is used to charge the supercapacitor ($S_{cap}$) again.

\par




\section{Proposed Methods for Improving Reliability  with proper Detection and Mitigation}
\label{sec:AnalogTrojandetection}

In our proposed EHS, different modules like RO and counter in FSM are the primary source of reliability degradation. The reliability is explored in terms of ripple at the output, intentional aging that affects RO's oscillation frequency, and a specific Trojan that may cause a premature reset of the counter in the MPPT unit. These reliability issues and suitable mitigation techniques are briefed in this section.

\subsection{Factors Influencing Ripples and its Mitigation}

\subsubsection{Ripple Analysis}

In a MOSFET the trail from source to drain is diagrammatic by a linear resistance adequate to R ($R_{on}$) as in Eqn. \ref{eq:Onresistance}: 
\begin{equation}
\label{eq:Onresistance}
{R_{on}} = \left( \frac{1}{{{\mu _n}{C_{ox}}\frac{W}{L}\left( {{V_{GS}} - {V_T}} \right)}} \right).
\end{equation}

\begin{figure}[htbp]
	\centering
\includegraphics[width=0.70\textwidth]{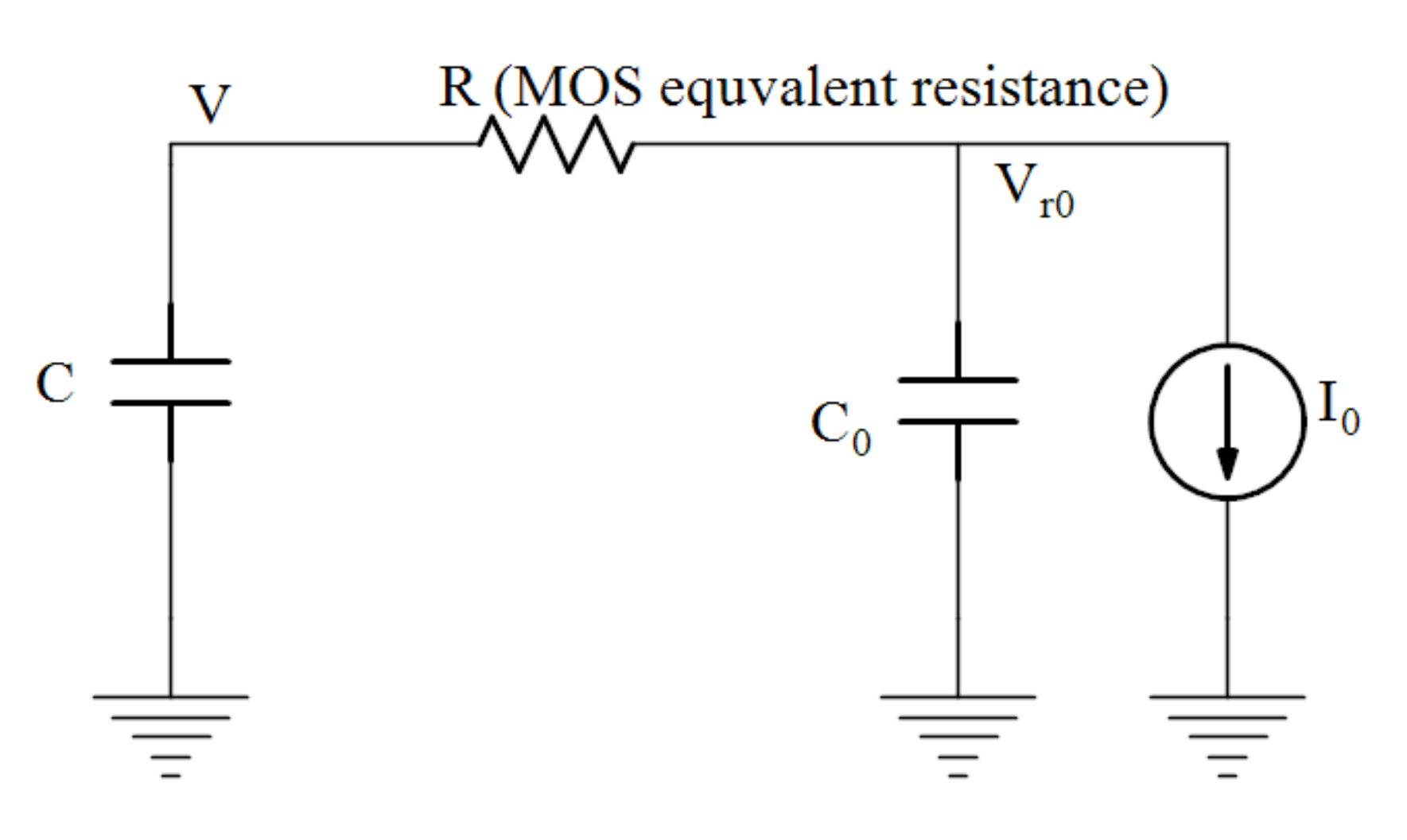}
\caption{Simplified equivalent circuit at output stage of Charge Pump.}
\label{fig:MOSMODEL}
\end{figure}

In the charge pump, the charge transfers from one capacitor to another occur through MOS switches with required clock frequencies for switching \cite{ jaw2012analysis}. The voltage transfer in charge pumps can be explained as depicted in Fig. \ref{fig:MOSMODEL}.

Since the charge transfer from one capacitor to another has to occur in time interval $T/2$ requires some time to pass through the MOS equivalent resistance $R$. Thus $V_{r0}$ increases from zero to ${V_{r0,\max }}$ , which is the ripple and return to 0.
The ${C_0}$ is in parallel with C. ${R{C_p}}$ is much less than $T/2$. ${V_{r0}}$ can be approximated as Eqn. \ref{eq:APPROX}.
\begin{equation}
\label{eq:APPROX}
{V_{r0}} = \frac{{T{I_0}}}{{C + {C_0}}}\left( {1 - {{\exp }^{ - \frac{t}{{R{C_p}}}}}} \right) - \frac{{{I_0}t}}{{C + {C_0}}}
\end{equation}

The ${V_{r0}}$ is assumed to achieve its maximum value ${V_{r0,\max }}$, at time ${t_{\max }}$, then ${V_{r0,\max }}$ can be obtained by solving 
$d {V_{r0(t=tmax)}}$/$d t$ = 0. The corresponding ${t_{\max }}$ and ${V_{r0, \max }}$ are as follows: 
\begin{eqnarray}
\label{eq:TIMEMAX}
{t_{\max }} & = & R{C_p}\ln \left( {\frac{{T/2}}{{R{C_p}\left( {1 - \exp \left( { - T/2R{C_p}} \right)} \right)}}} \right) .\\
%
\label{eq:RIPPLEMAX}
{V_{r0,\max }} & = & \frac{{T{I_0}}}{{2\left( {C + {C_0}} \right)}} - \frac{{R{I_0}C{C_0}}}{{{{\left( {C + {C_0}} \right)}^2}}}\left( {1 + \ln \left( {\frac{{T/2}}{{R{C_p}}}} \right)} \right) .
\end{eqnarray}

From Eqn. \ref{eq:RIPPLEMAX}, the following are evident:
 \begin{description}
\item[$\bullet$ ] The ripples at the output voltage decrease if $T$ (1/$f_{ossc}$) decreases or $I_{0}$ is small or by taking a more significant value of the load capacitor. The $W/L$ and load capacitor, if not appropriately chosen, then there is a degradation in boosting.
\end{description}

From the above discussion, it is observed that by making $T$ constant, i.e., the RO's oscillation frequency is made constant. If the oscillation frequency degrades from its actual value, then the $T$ will change, and ultimately ripples at output increase as mentioned in Eqn. \ref{eq:RIPPLEMAX}.

\subsubsection{Reliability Degradation}

As the ripple depends on $f_{ossc}$ of RO, the solution lies in designing RO with constant $f_{ossc}$. As the $f_{ossc}$ is threshold voltage-dependent, any change in MOS threshold voltage affects the ripple, causing reliability degradation \cite{rahman2016aging}. The
threshold voltage ($V_{th}$) variation is either reversible (due to the temperature change) or irreversible (due to aging). Aging mechanisms introduce a permanent ripple at the output, whereas ripple due to temperature variation is reversible.

This section briefs how aging affects the oscillation frequency of RO and its mitigation. As discussed in \cite{sahoo2018novel}, out of different sources of aging like NBTI (negative bias temperature instability) and HCI (hot carrier injection), NBTI is the primary cause of degradation in the oscillation frequency of RO. The degradation in $f_{ossc}$ of RO due to NBTI occurs even if the RO is in an idle state. The impact of NBTI on conventional RO and aging tolerant RO proposed in \cite{sahoo2018novel} is shown in Fig. \ref{fig:NBTI}.

\subsubsection{ Aging Tolerant Ring Oscillator for Mitigation }

\begin{figure}[t]
\centering
\includegraphics[width=0.75\textwidth]{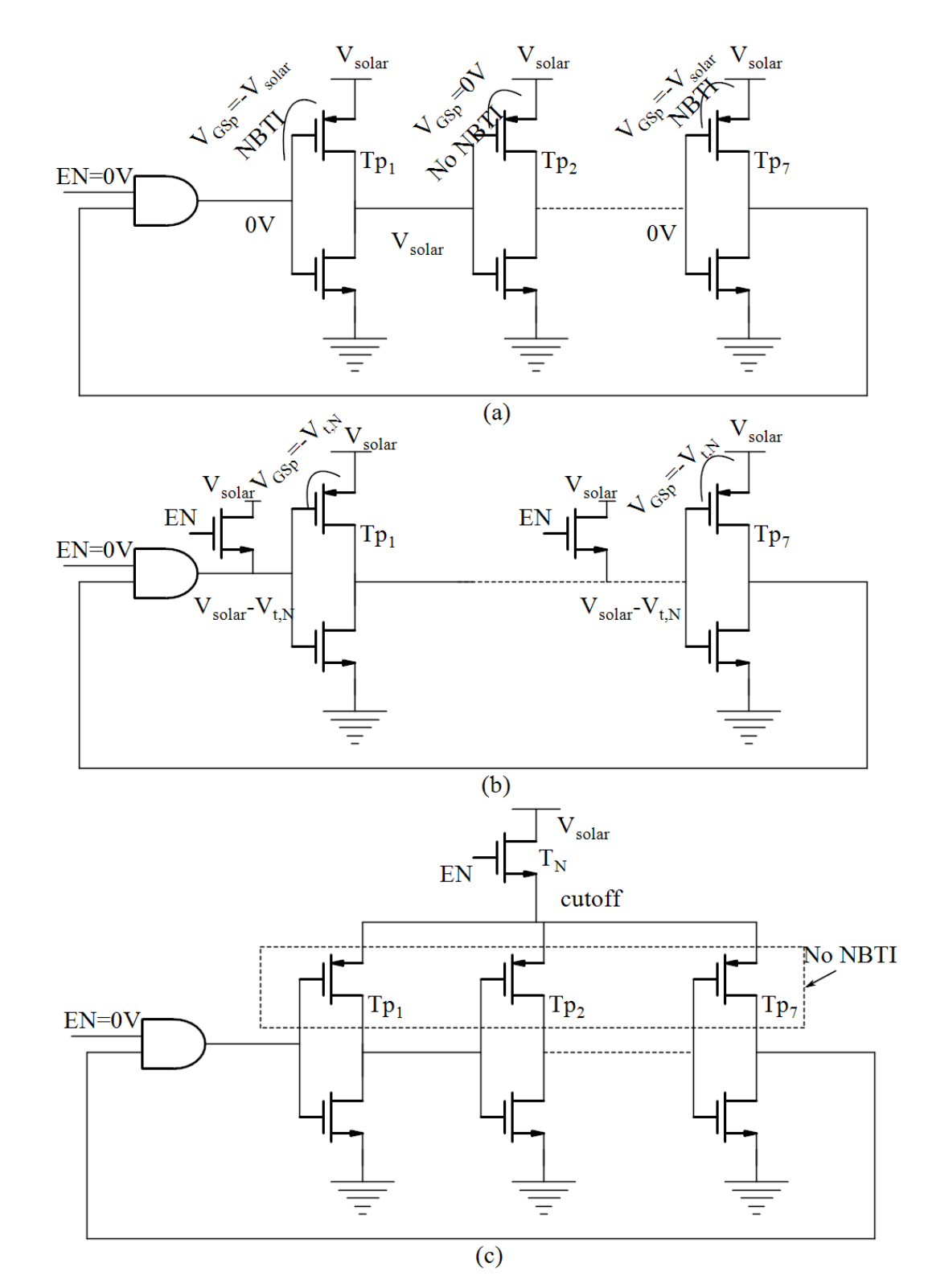}
\caption{NBTI effect (a) CMOS RO (b) ARO (c) Reduced $V_{solar}$ ARO.}
\label{fig:NBTI}
\end{figure}

The impact of aging on RO is observed by considering conventional RO, ARO \cite{rahman2016aging}, and the recently proposed aging tolerant RO designed using reduced supply voltage \cite{sahoo2018novel} as depicted in Fig. \ref{fig:NBTI}.
\par
The impact of NBTI becomes more pronounced during the non-oscillation mode of RO, as shown in Fig. \ref{fig:NBTI}. In conventional RO (Fig. \ref{fig:NBTI} (a)), half of the PMOS is always undergoing NBTI stress due to logic-0 at the gate terminal $V_{GS} = - V_{solar}$. However, in ARO, the impact of NBTI is reduced by lowering the magnitude of negative bias from $V_{solar}$ to $V_{tn}$. Although the degradation due to NBTI is less in ARO than conventional RO, it still suffers from NBTI due to lower negative bias across all the PMOS in the cascaded inverter. The architecture proposed in \cite{rahman2016aging}, eliminates the impact of NBTI, as shown in Fig. \ref{fig:NBTI} (c). By removing the supply voltage entirely during the non-oscillation mode (EN = 0), all the PMOS remains free from NBTI stress.
\par
Similarly, the degradation in threshold voltage due to HCI is observed during the oscillation mode of RO. The switching action at the gate of NMOS determines the rate of degradation. All the three considered RO oscillates at the same frequency hence the impact of HCI remains the same.

The interpretation from the above discussion is, the impact of NBTI is the prime cause of degradation in the oscillation frequency. As the RO with reduced supply voltage experiences a lower NBTI effect, it undergoes a lower deterioration rate in oscillation frequency.

 \begin{figure}[t]
\centering
\subfloat[Degradation of Oscillation frequency with time]{\label{fig:DEGRADATION}\includegraphics[width=0.650\textwidth]{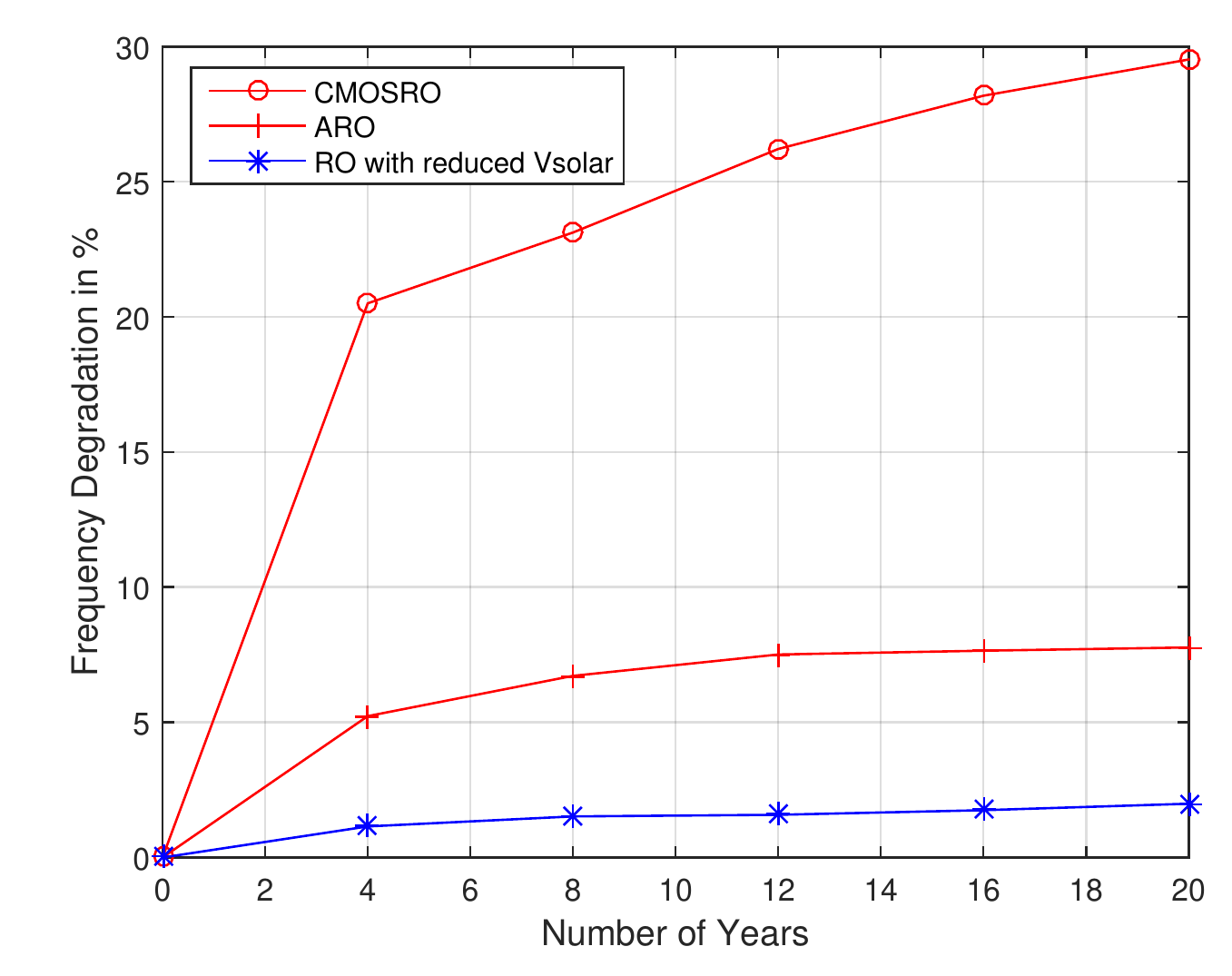}}\\
\subfloat[Ripples at output of Charge Pump (in mV) v/s time]{\label{fig:DEGRADATION1}\includegraphics[width=0.650\textwidth]{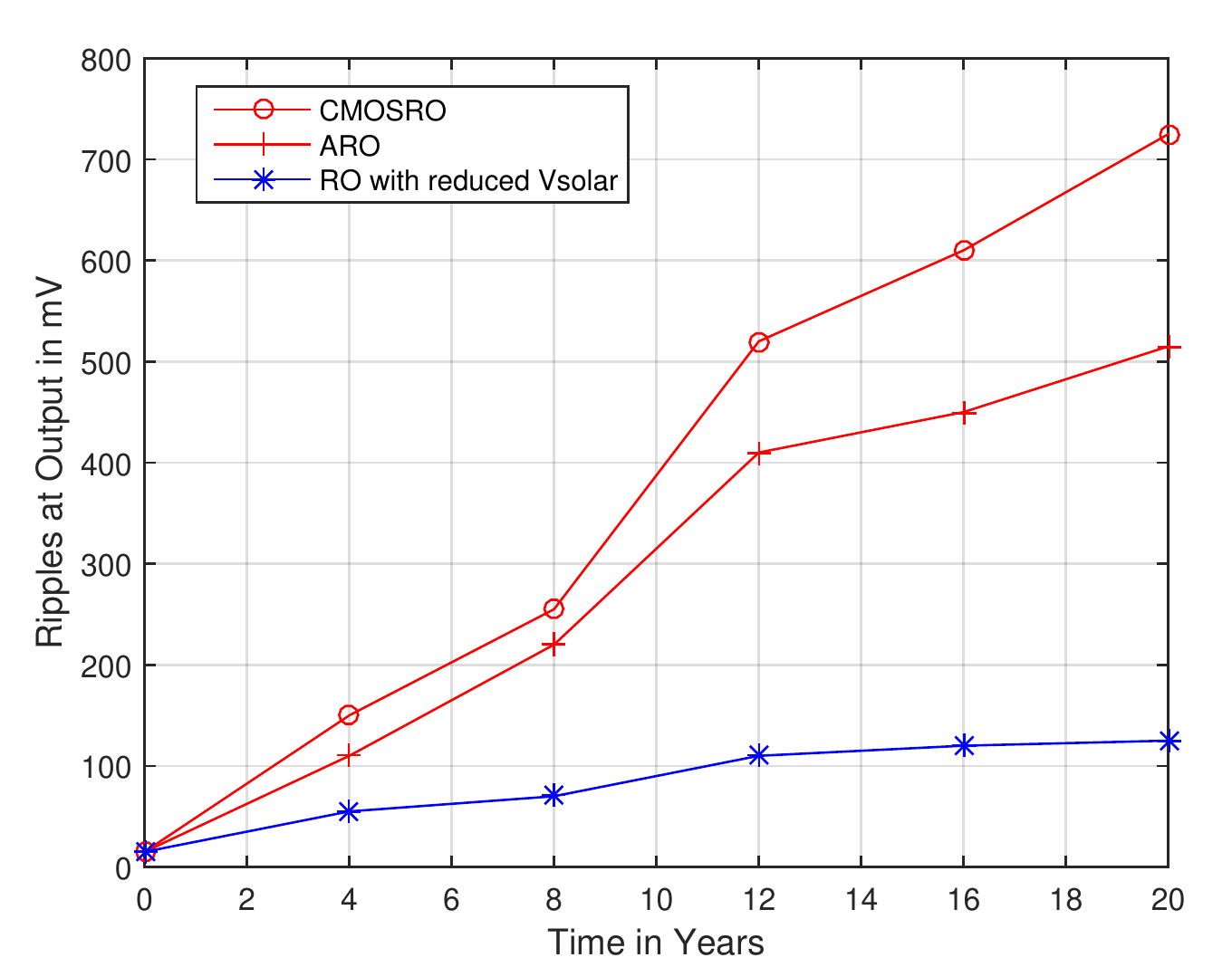}}\\
\caption{Degradation in Oscillation Frequency causing more ripples at Converter output.}
\end{figure}

Fig. \ref{fig:DEGRADATION}, and Fig. \ref{fig:DEGRADATION1} depicts the rate of degradation in the oscillation frequency of RO due to aging for 4, 8, 12, 16, and 20 years. The result shows, the RO in \cite{sahoo2018novel} experience a lower rate of degradation (3\%) as compared to conventional CMOS RO (30\%) and ARO (10\%). The lower rate of degradation in the RO shows that the oscillation frequency is less influenced by aging and remains constant over a more considerable period. The less degradation of frequency results in fewer ripples at the output. Further, the adversary may cause intentional aging to affect the $f_{ossc}$ of RO; thus, the reliability is also affected. As the RO with reduced $V_{solar}$ is aging tolerant, intentional aging due to temperature variation does not reduce reliability.

\subsection{ Trojan Influence on EHS and Its Detection}

\begin{figure}[t]
\centering
\includegraphics[width=0.55\textwidth]{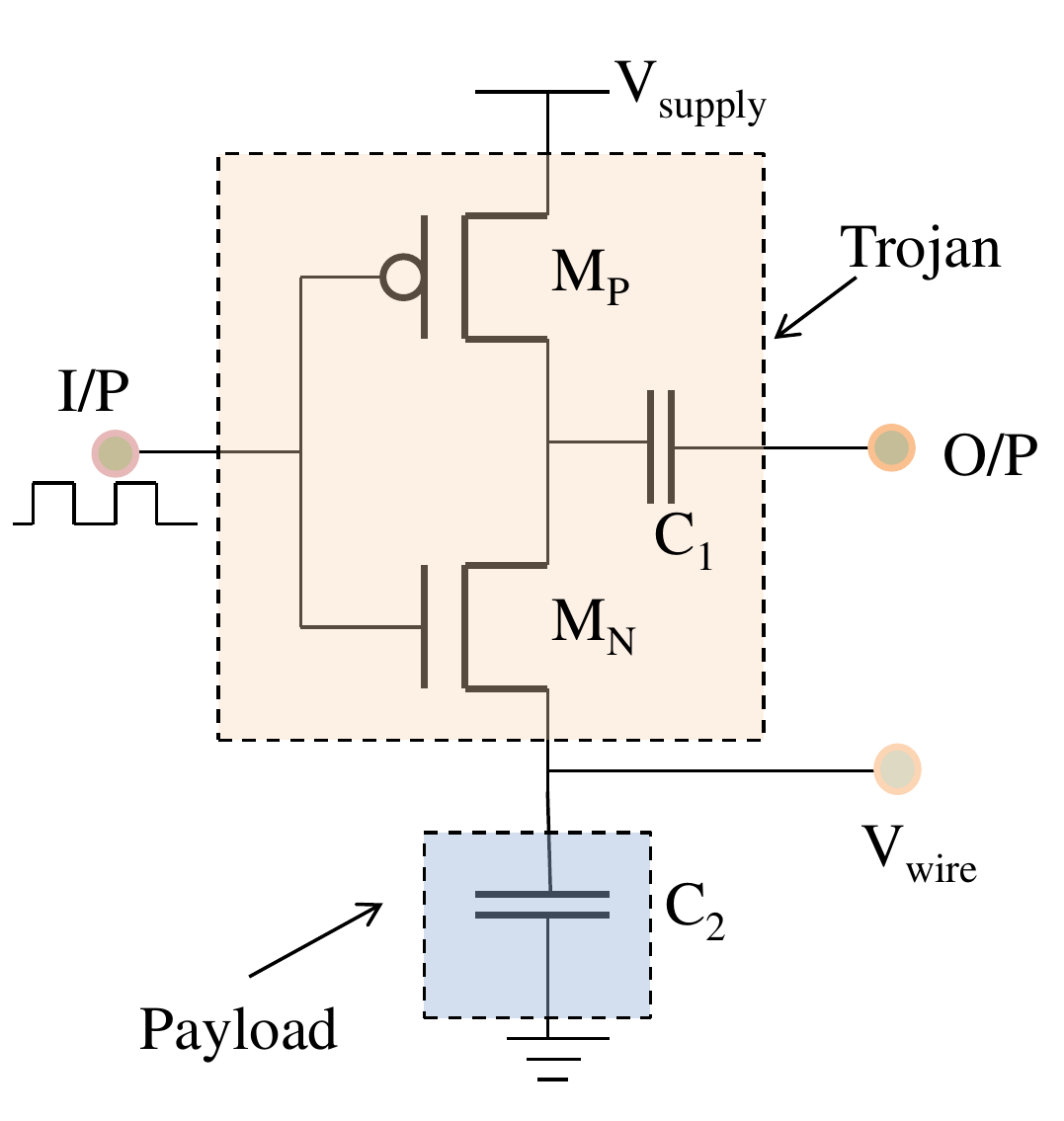}
\caption{Schematic Diagram of a Analog Malicious Trojan (A2) Hardware.}
\label{fig:A2Trojan}
\end{figure}

Although varied Trojan attacks are possible, the projected design is additional sensitive to A2-based Trojan because of the MPPT module's counter. A brief description of Trojan and the corresponding aging mitigation is discussed. In this design, we tend to used a particular Analog Hardware Trojans (A2) that uses capacitors, that charges from the close values in wires as they transit between the digital logic values. When the capacitor gets fully charged, it affects the performance of the actual circuit. In our design, the FSM controls the total operation of the MPPT. The FSM uses a counter and additional circuitry for assigning different time slots for issuing various control signals. The counter, if reset intentionally using Trojan, then the MPPT operation gets interrupted. It will affect the EHS performance drastically. Fig. \ref{fig:A2Trojan} depicts the circuit diagram of A2 Trojan based on charge sharing and capacitative coupling.

\begin{figure}[htbp]
\centering\includegraphics[width=0.85\textwidth]{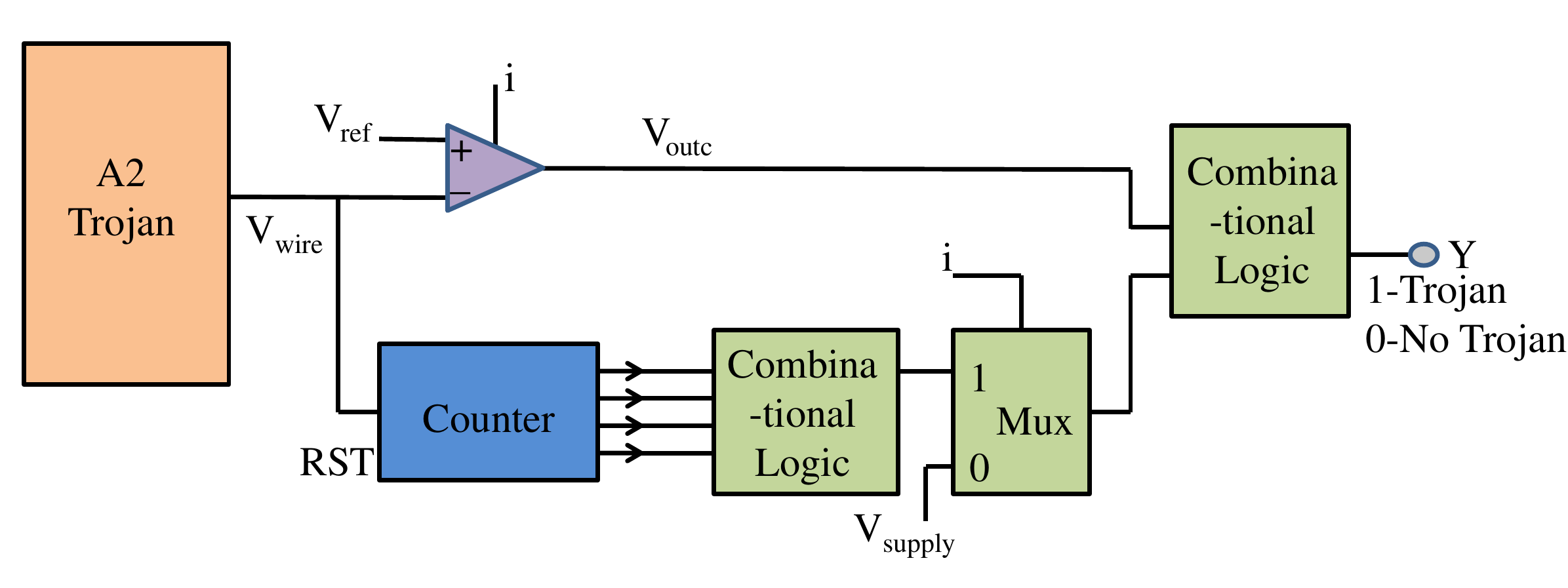}
\caption{Proposed Detection Mechanism for A2 Trojan.}
\label{fig:A2Trojandetection}
\end{figure}

The attacker tries to design an A2 Trojan that forcefully drives the counter's output to zero before completing the required 16-MPPT cycles.
The attacker decides a toggling frequency of inverter input and decides the capacitance value; so that voltage at the RST (reset) terminal is gradually build up. When the voltage exceeds a certain threshold, it drives the RST to zero.

\subsection{A2 Trojan Detection circuit}

Our objective is to check any voltage built up on the RST node during counting operation. The detection mechanism and associated methodology is shown in Fig. \ref{fig:A2Trojandetection}.

The Trojan is used to charge the payload, which triggers the counter's reset input, thereby affecting the MPPT procedure before its completion. In our EHS, one MPPT cycle consists of 32 clock cycles (6.66$\mu$S X 32=231.12$\mu$S). The CVM process for MPPT achievement has a maximum of 16 thermometer bits ($b_{0}$-$b_{15}$) that takes 16 MPPT cycles (213.12$\mu$S X 16=3410$\mu$S). The signal $i$ is high for 3410$\mu$S to detect the presence of a Trojan. The total procedure for the detection of A2 Trojan is explained below.

\begin{description}
  \item[$\bullet$] The reset ($RST$)=0, counter is in normal operation mode and if 1, counter will be reset.
  \item[$\bullet$] The comparator compares (when $i$=1) the output $V_{wire}$ with $V_{ref}$, if $V_{wire}$ $>$ $V_{ref}$ then $V_{outc}$=0 else 1 (first stage of Trojan detection).
  \item[$\bullet$] The counter outputs are normally non-zero during the MPPT cycle. The output becomes zero during MPPT if there is a Trojan.
  \item[$\bullet$] The enable bit during MPPT period ($i$) is high for 3140 $\mu$S for enabling the Trojan detection circuits. 
  \item[$\bullet$] The counter outputs passed to the logic circuit for generating a combined output.
  \item[$\bullet$] The output of the logic block is given as input to the multiplexer having selection line as $i$. The $i$ if high the combined output of the logic block is passed else the output is high.
  \item[$\bullet$] The high in the multiplexer output indicates no Trojan and low indicates there is Trojan during $i$=1.
  \item[$\bullet$] The comparator and multiplexer output finally passed to the logic block which detects a Trojan (if 1) and no Trojan (if 0). 
\end{description}


\section{Experimental Results}
\label{sec:Experimentalresults}

\subsection{ Experimental Setup }

The reliable Trojan resilient PV-EHS is designed in CMOS 90nm technology library. 
The capacitors used in the voltage tripler and capacitor banks are MIM (Metal-Insulator-Metal) capacitors. The digital capacitor bank is employed for impedance matching through CVM. 
The solar voltage is within the range of 1-1.5V (with temperature 27$^\circ$C). The load was designated with a resistor in range 200K$\Omega$ to 10M$\Omega$ in parallel with a supercapacitor having 33mF value.
\par
The RelXpert simulator in the virtuoso environment is used for aging analysis. It uses the model library given by the foundry, which supports both NBTI and HCI effects. Both fresh (at $t$ = 0) and aged SPICE netlist are extracted at different aging intervals ($t$ = 4, 8, 12, 16, 20 years) to measure the ripples at the output. The ripple variation due to aging is observed by considering both conventional RO and aging tolerant RO (with reduced $V_{solar}$).

\subsection{EHS Simulation with MPPT }

\begin{figure}[htbp]
\centering
\subfloat[Clocks and Bias generation using RO, NOCG, ACP, and LS]{\label{fig:MPPTFINAL}\includegraphics[width=0.75\textwidth]{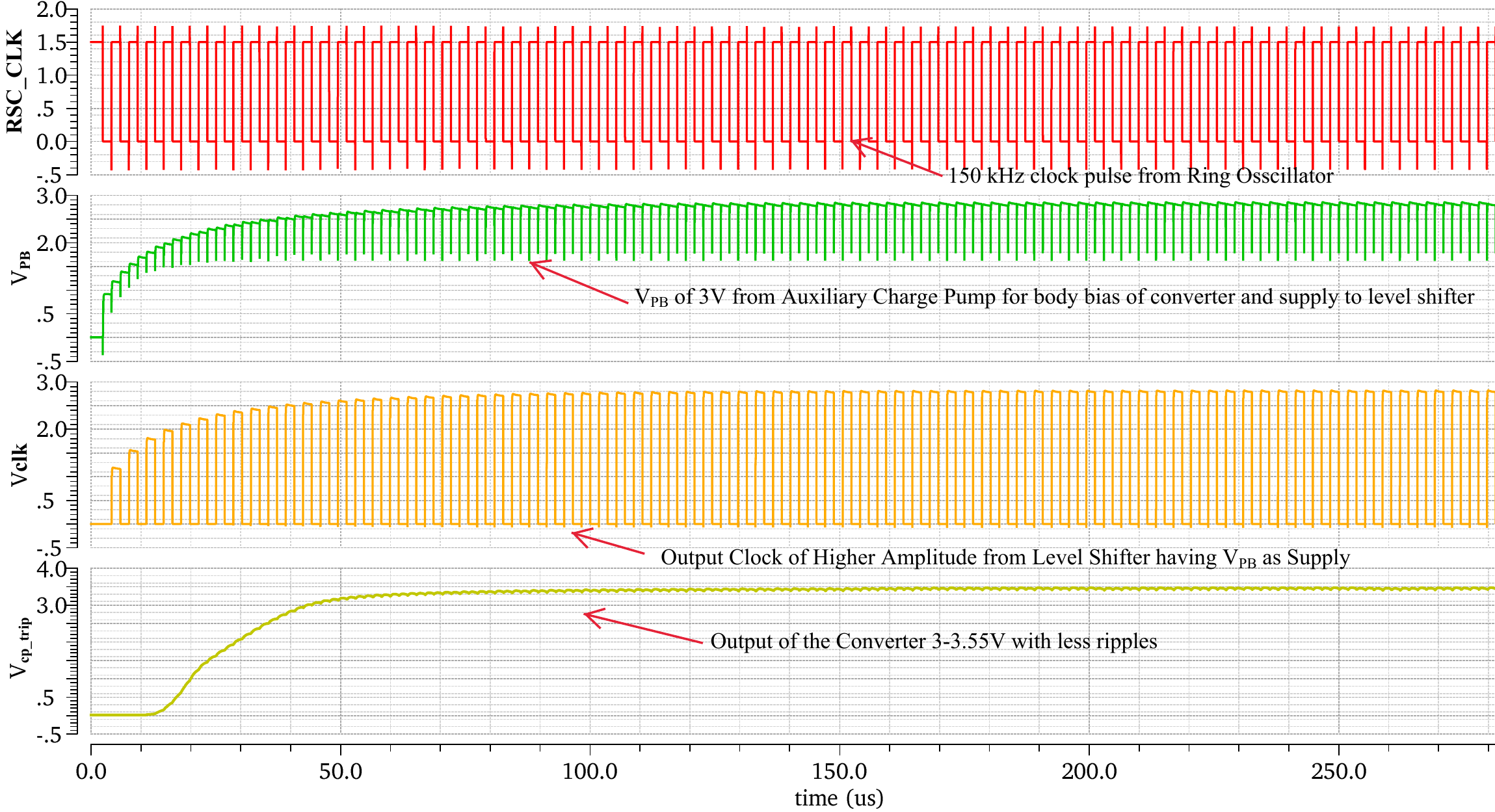}}\\
\subfloat[Control signals related to FSM for MPPT]{\label{fig:MPPTFINAL1}\includegraphics[width=0.75\textwidth]{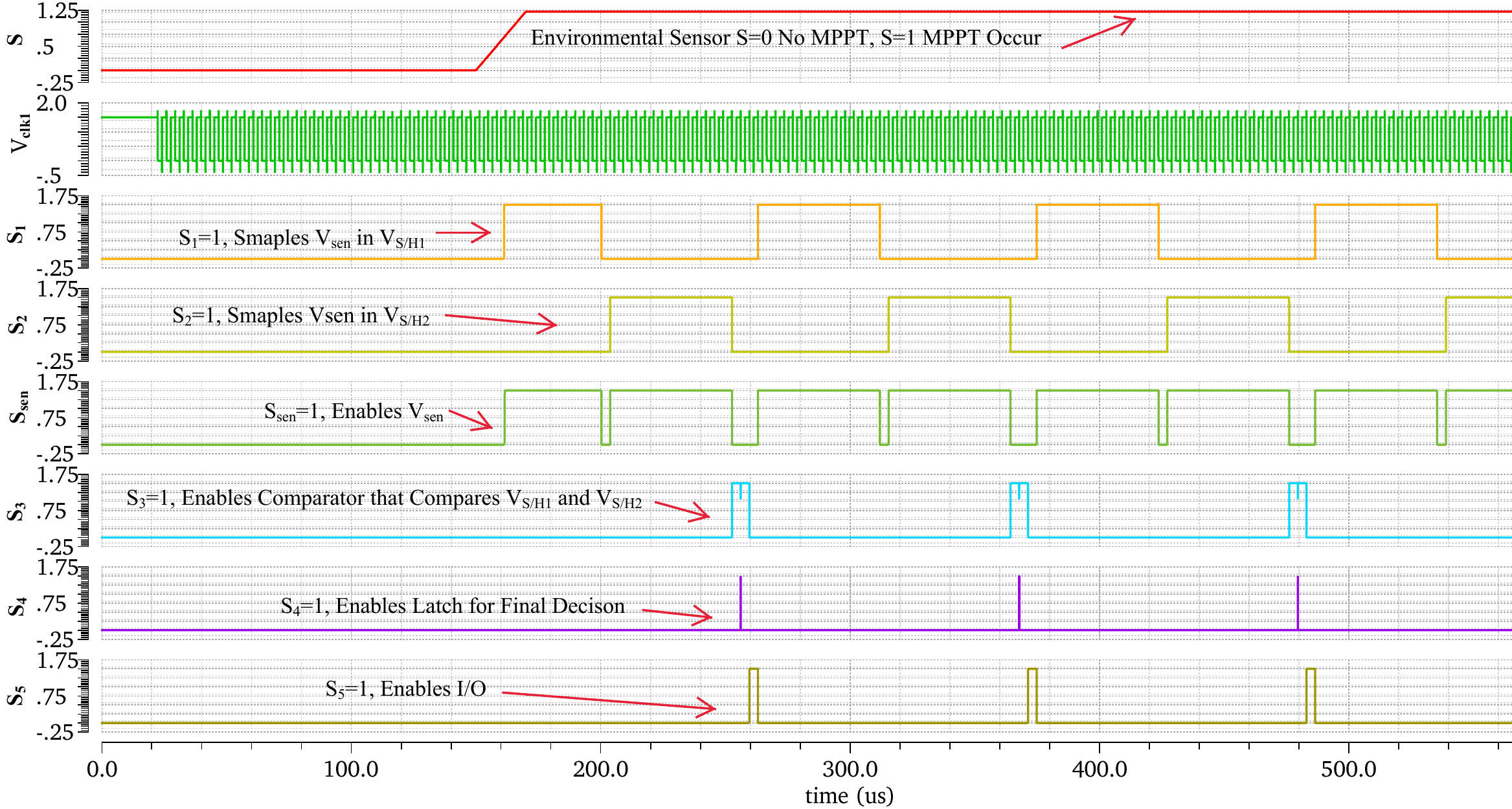}}\\
\subfloat[Simulation result of EHS  including control signals with MPPT achievement]{\label{fig:MPPTFINAL2}\includegraphics[width=0.75\textwidth]{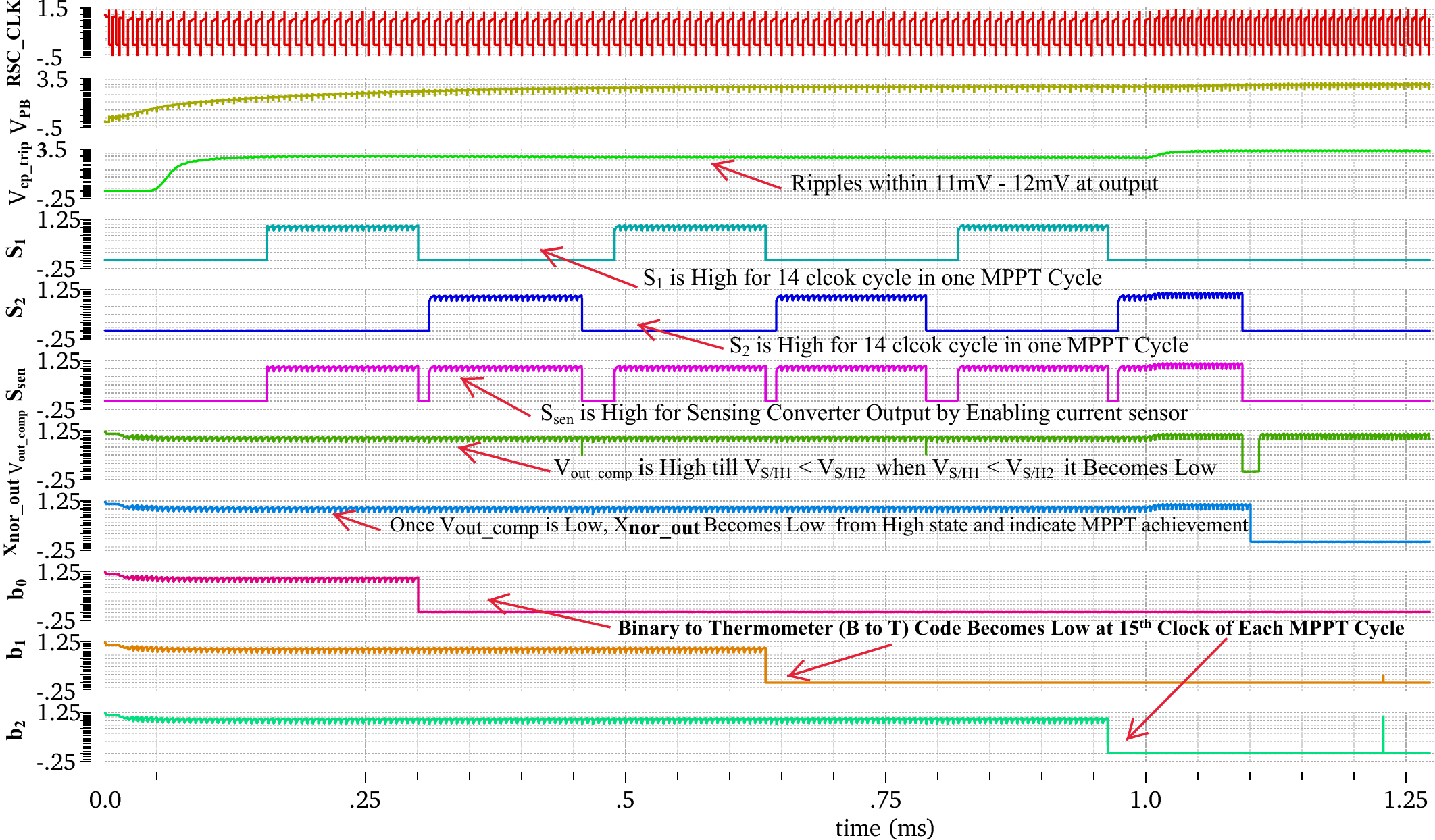}}
\caption{Different Bias and Control Signals for converter and MPPT in Harvesting System.}
\end{figure}

The generation of clocks, higher bias voltage and boosted clock signals (like $V_{clk1}$, $V_{clk2}$, $V_{PB}$, and, $V_{clk}$) for self-sustainable operation of entire system is shown in Fig. \ref{fig:MPPTFINAL}.
\begin{description}
\item[$\bullet$ ] 
Fig. \ref{fig:MPPTFINAL} depicts the clock output ($RSC\_CLK$) generated by the aging tolerant RO ($f$ = 150 KHz).
   The higher bias voltage ($V_{PB}$=3V) generated is shown in Fig.\ref{fig:MPPTFINAL} and is used by the level shifter (LS) as a supply voltage and for DC-DC converter as a body bias (for PMOS devices). 
  Fig.\ref{fig:MPPTFINAL} shows the corresponding boosted clock $V_{clk}$ from LS and is fed to the DC-DC converter.
\end{description}

\par
Fig. \ref{fig:MPPTFINAL1} depicts the simulation result of a complete MPPT cycle with 32 clocks along with the associated signals generated by the FSM. As per design consideration, when the solar voltage $V_{solar}$ falls below the assumed threshold, an environmental sensor (S) triggers MPPT.  
\begin{description}
    \item[$\bullet$ ] The generated control signals from the FSM ($S_{1}$, $S_{2}$, $S_{sen}$, $S_{3}$, $S_{4}$, $S_{5}$) after triggered by the environmental sensor (S) is shown in the Fig. \ref{fig:MPPTFINAL1}.
    As depicted in Fig. \ref{fig:MPPTFINAL1}, S remains Low (0 V) for 150$\mu$S for normal operation and switched to High for MPPT initialization.
    \item[$\bullet$ ] After MPPT initialization, signals $S_{1}$, $S_{2}$ are high for 14-clock cycles as shown in Fig. \ref{fig:MPPTFINAL1} for sensing power information $P_{n}$ and $P_{n+1}$.
   Fig. \ref{fig:MPPTFINAL1} depicts $S_{sen}$ remains High when $S_{1}$ or $S_{2}$ is High and is Low in between ($S_{1}$ and $S_{2}$ for one cycle) for CVM and last two periods during final decision for MPPT. 
    \item[$\bullet$ ] As depicted in Fig. \ref{fig:MPPTFINAL1}, control signal $S_{3}$ is high ($30^{th}$, and $31^{st}$ cycles) for comparing $P_{n}$ and $P_{n+1}$, and \: $S_{4}$ is High (after $30^{th}$ cycle) for a small duration, for decision making about MPPT achievement, whereas $S_{5}$ initiates I/O operation (if any at the end of $31^{st}$ cycle). 
\end{description}

\begin{figure}[t]
\centering
\subfloat[Power consumption (in \%) by each unit in the EHS]{\label{fig:Powerconsumption_DIE}\includegraphics[width=0.70\textwidth]{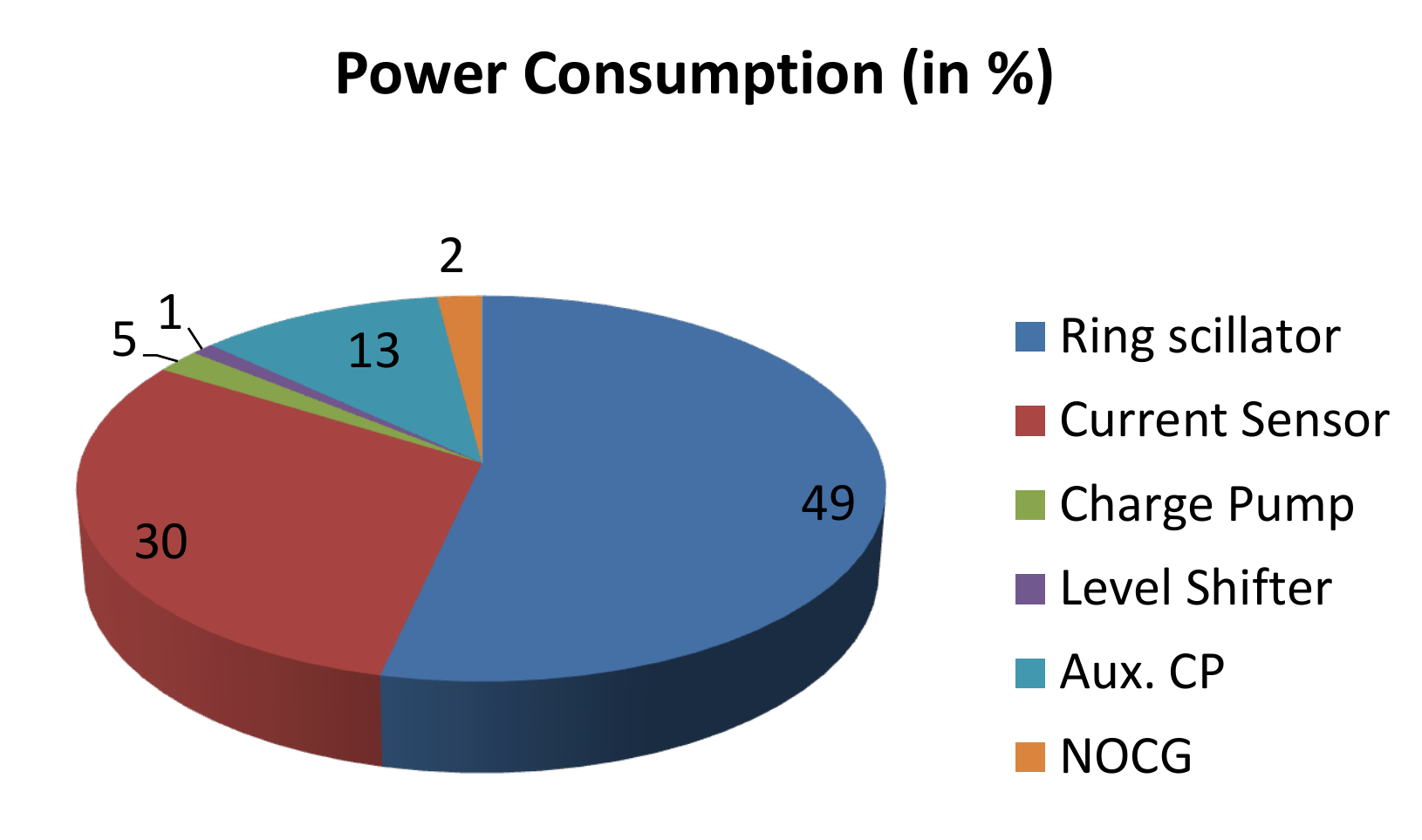}}\\
\subfloat[DIE Photograph of EHS Chip given for Fabrication]{\label{fig:Powerconsumption_DIE1}\includegraphics[width=0.70\textwidth]{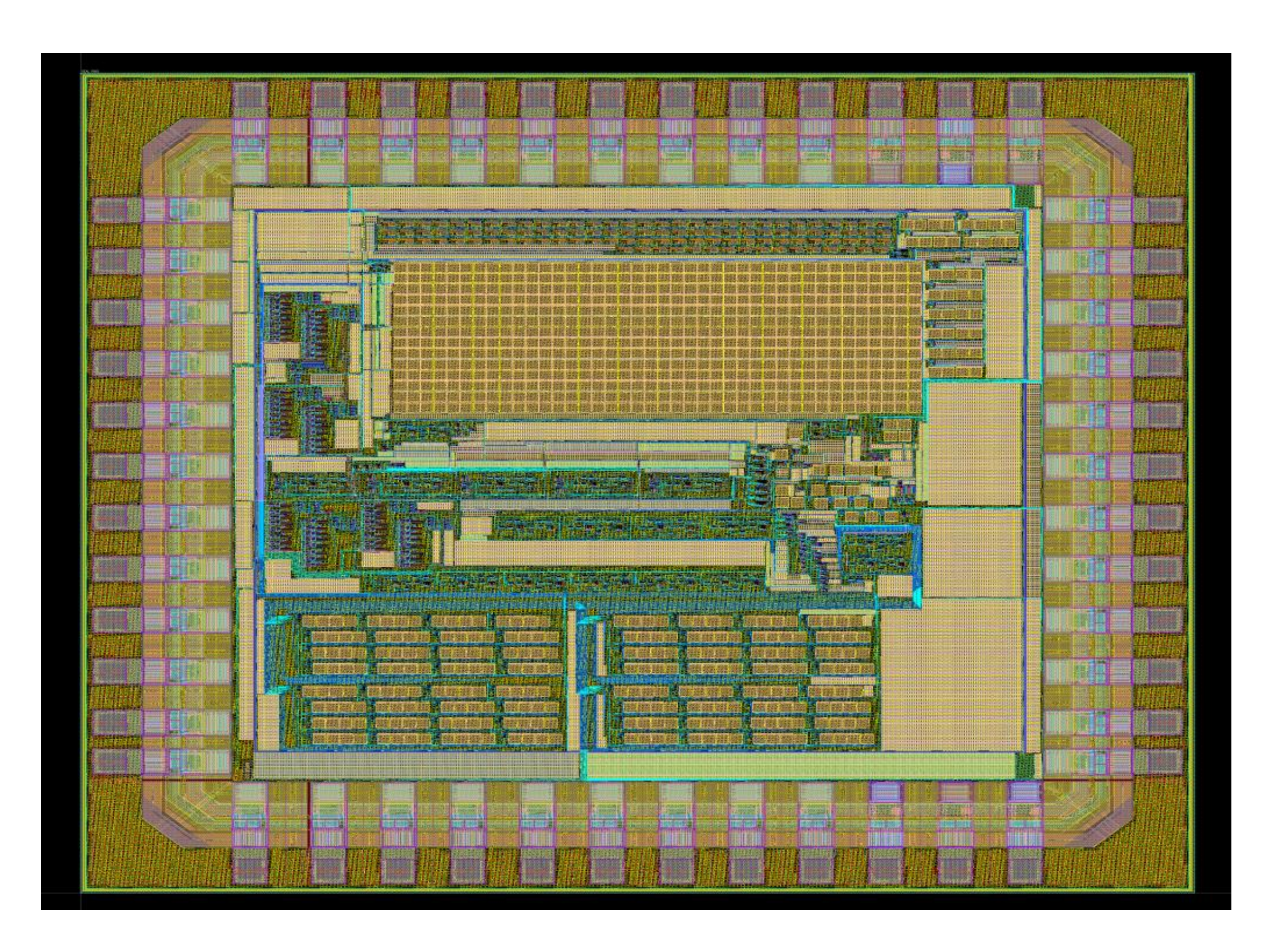}}\\
\caption{Power Consumption by Each Module and Die Photograph of EHS.}
\end{figure}

The number of MPPT cycles needed to achieve the MPP along with the signals needed for CVM during MPPT process is shown in Fig. \ref{fig:MPPTFINAL2}. 
The change in control signals and decision making process after MPPT achievement is presented. 
For experimental validation, the value of $V_{solar}$ ($V_{solar}$=1-1.5V) is varied around the threshold voltage. 
After 150$\mu$S, it is assumed that the $V_{solar}$ comes below the required level and the associated signals are shown in Fig. \ref{fig:MPPTFINAL2}.
As depicted in Fig. \ref{fig:MPPTFINAL2}, the output voltage ($V_{cp\_trip}$) of the charge pump, is in the range of 3-3.55 V.
The control signals $S_{1}$ and $S_{2}$  is used to sample and hold the power information as $V_{out1}$ and $V_{out2}$ respectively from $V_{sens}$.

The programming of digital capacitor banks connected to the power capacitor $C_{u}$ of the converter is carried out using thermometer codes ($b_{0}$ to $b_{15}$), generated by the FSM. The sequence of operation for MPPT is discussed as follows:
\begin{description}
 \item[$\bullet$ ] As depicted in Fig. \ref{fig:MPPTFINAL2}, the thermometer codes ($b_{0}$, $b_{1}$, $b_{3}$ ....) are initially High (for the full load operation). A transition from High to Low indicates ($15^{th}$ clock of each MPPT period) CVM.
\item[$\bullet$ ] The comparator output ($V_{out\_comp}$) remains High until $P_{n}$ is less than $P_{n+1}$. When $P_{n}$ is higher than $P_{n+1}$, the output of the comparator goes from High to Low, which indicates about MPPT achievement, which is shown in green line in Fig. \ref{fig:MPPTFINAL2}. 
\item[$\bullet$ ] Once the $V_{out\_comp}$ switches from High to Low, the XNOR gate output ($X_{nor\_out}$), which is initially high switched to Low, this indicates about MPPT achievement as shown in Fig. \ref{fig:MPPTFINAL2}.
\end{description}

The simulation result in Fig. \ref{fig:MPPTFINAL2} confirms that once the MPP is achieved, the MPP is locked. In this experimental evaluation, after 3 MPPT cycles, the MPP is achieved ($P_{n}$ $>$ $P_{n+1}$). Once MPP is locked, all the control signals get reset by the FSM. The supercapacitor ($S_{cap}$) continues to charge, as $S_{sen}$ is Low. The above process for MPPT repeats, when the $V_{solar}$ drops due to change in environmental parameters.

\subsection{Reliability Analysis}

As briefed in section \ref{sec:ResilientIoT}, degradation in the oscillation frequency of RO due to aging causes more ripples at the output. Higher ripple is observed when the RO undergoes continual aging. This increase in ripple is due to a higher rate of degradation in the oscillation frequency of CMOS RO (as shown in Fig. \ref{fig:DEGRADATION}). The detail analysis of the degradation in frequency with time is presented in Table \ref{tab:FREQUENCYDEGRADATION}.

\begin{table}[htbp]
	\centering
	\caption{Degradation in $f_{ossc}$ with time}
	\label{tab:FREQUENCYDEGRADATION}
	{\begin{tabular}{ |c|c|c|c| } 
\hline
			Time in Years & Conventional RO & ARO & Proposed RO\\ 
			\hline
			\hline
			0 & 150 kHz & 150 kHz & 150 kHz \\ 
			\hline
			5 & 130 kHz & 139.9 kHz & 148.2 kHz \\ 
			\hline
			10 & 115.5 kHz & 138.7 kHz & 147.6 kHz \\ 
			\hline
			15 & 111 kHz & 138.5 kHz & 147.3 kHz \\ 
			\hline
	     	20 & 110 kHz & 138 kHz & 147 kHz \\ 
			\hline
	\end{tabular}}
\end{table}

\begin{figure}[t]
\centering
\subfloat[Ripples at output with different temperatures]{\label{fig:DIFFERENTTEMPRIPPLEARO}\includegraphics[width=0.60\textwidth]{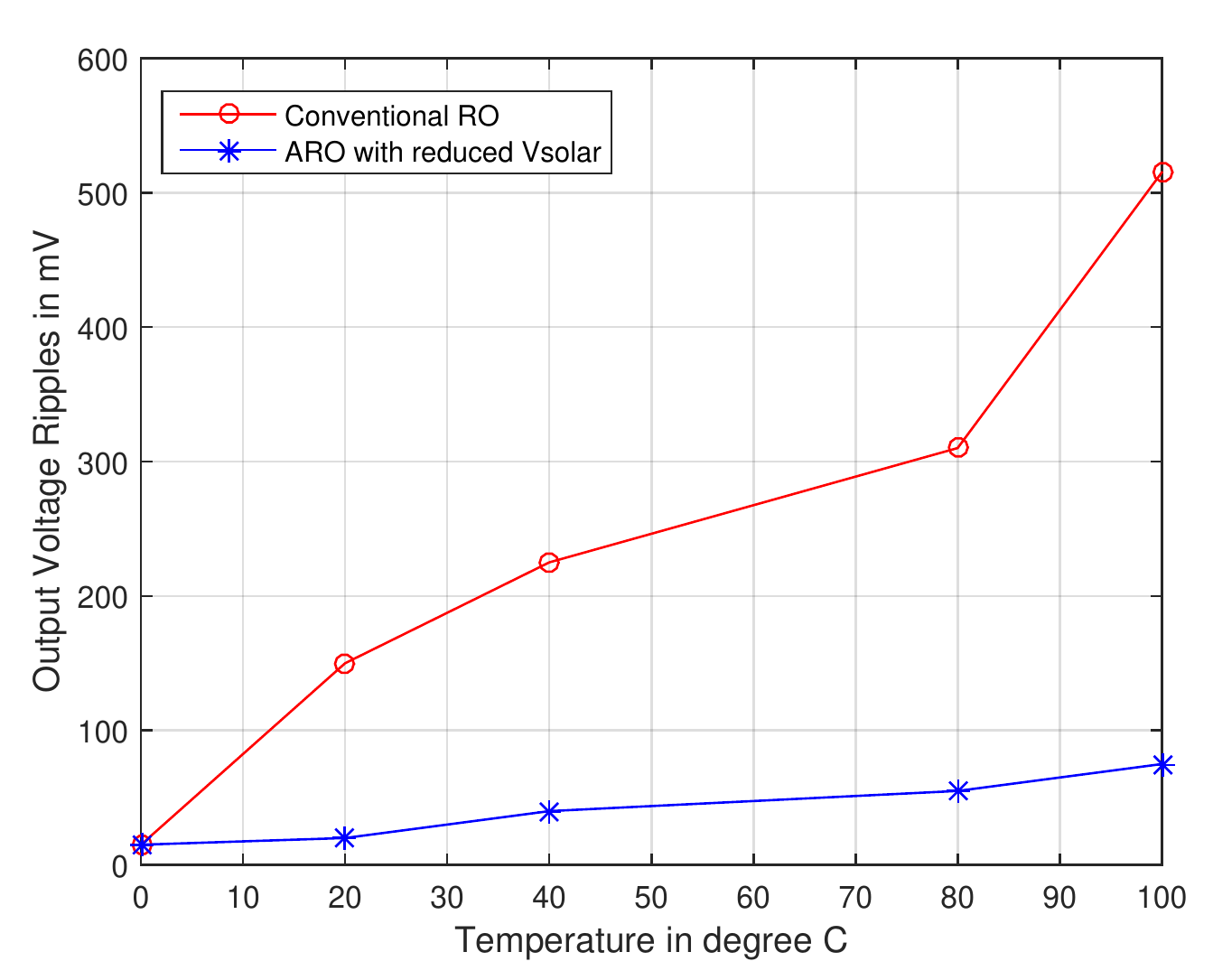}}\\
\subfloat[Ripples at output with ARO (reduced $V_{solar}$) versus time]{\label{fig:DIFFERENTTEMPRIPPLEARO1}\includegraphics[width=0.60\textwidth]{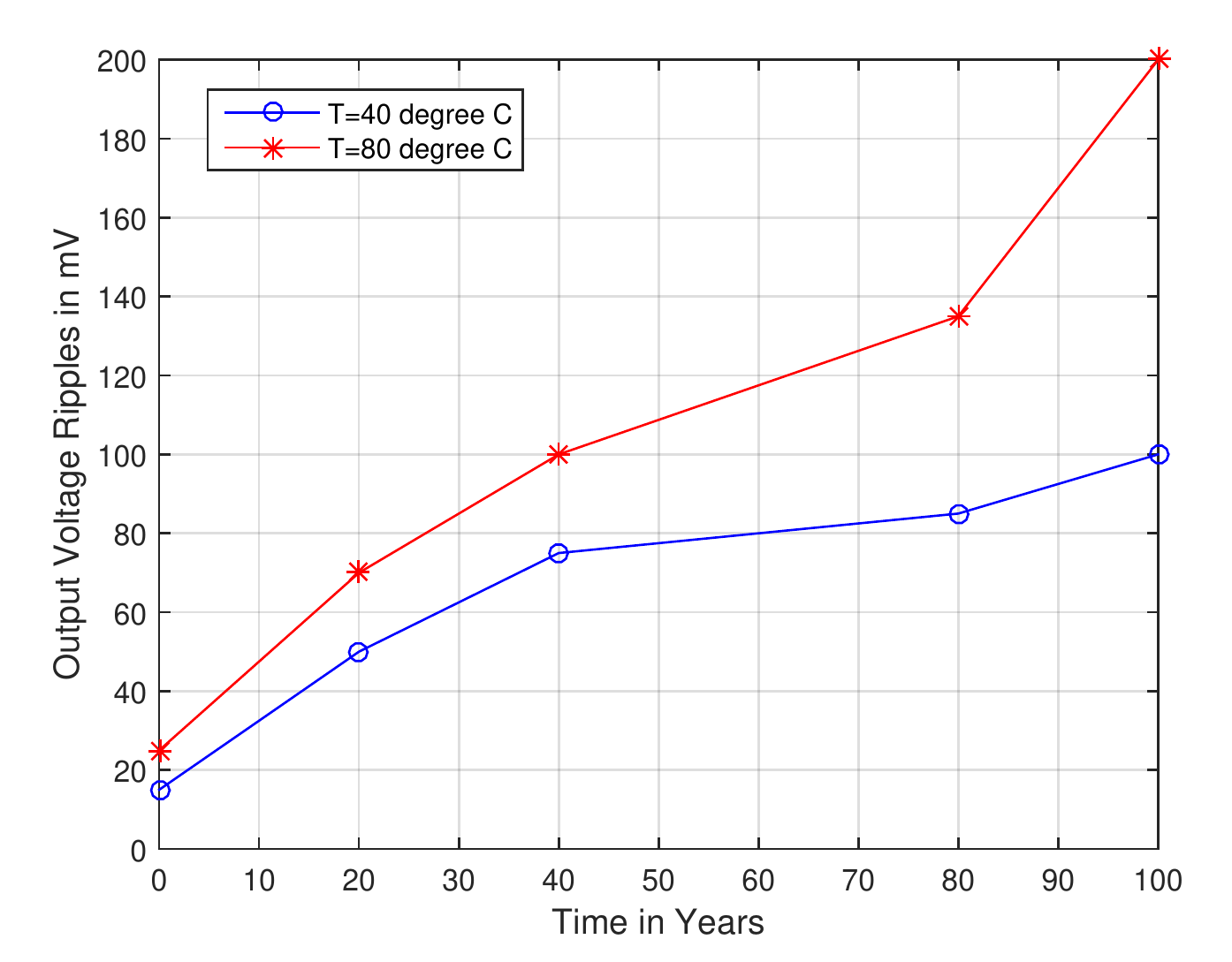}}\\
\caption{Ripple Analysis at Different Temperatures and with ARO (reduced $V_{solar}$).}
\end{figure}

To observe the ripples at the output, we have considered 3-different RO, i.e., CMOS RO, ARO \cite{rahman2016aging} and RO proposed in \cite{sahoo2018novel}. All the RO subjected to both NBTI and HCI stress continuously, and the ripples are observed at an aging interval of 4, 8, 12...20 years. Fig.\ref{fig:DEGRADATION1} shows the ripple due to aging. Among all the considered RO, a lower ripple is observed in the case of RO proposed in \cite{sahoo2018novel}, due to the lower rate of degradation in oscillation frequency (as shown Fig. \ref{fig:DEGRADATION}).
Further, the ripple variation due to the temperature rise is shown in Fig. \ref{fig:DIFFERENTTEMPRIPPLEARO}, for conventional CMOS RO, and RO proposed in \cite{sahoo2018novel}. Due to the lower temperature sensitivity of the RO presented in \cite{sahoo2018novel}, it exhibits lower ripples at a higher temperature than conventional CMOS RO.

Finally, the impact of aging is accelerated at a higher temperature. For analysis, two different temperatures, i.e., 40$^\circ$C and 80$^\circ$C are considered, and the ripples due to aging in the case of RO \cite{sahoo2018novel}, is shown in Fig.\ref{fig:DIFFERENTTEMPRIPPLEARO1}. The result shows a lower variation in ripple at a higher temperature.

\begin{table*}[htbp]
	\centering
	\caption{Comparison of different low energy solar harvesting systems.}
	\label{tab:STATEOFART}
	\begin{tabular}
{|p{2.0cm}|p{0.9cm}|p{0.9cm}|p{1.00cm}|p{0.85cm}|p{0.85cm}|p{1.5cm}|p{1.4cm}|p{1.4cm}|p{1.4cm}|} 
		\hline	
\textbf{Research Works} & \multicolumn{9}{c|}{\textbf{Feature or  Characteristics}}\\\cline{2-10}
		
		& Technology & Fully Integrated & Self-Sustaining & Input Range (V) & Output Range (V) & Power Throughput ($\mu$W) & Aging Sensor (For Counterfeiting IC) & Security Features Incorporated & Reliable (Aging Tolerant+ Trojan Detection mechanism)   \\ 
		\hline \hline
		Shao, et al. \cite{shao2009design} & 350nm & Yes & No & 2.1-3.5 & 3.6-4.4 & 100-775 & No & No & No \\
		\hline
		
		Kim, et al.	\cite{kim2011regulated} & 350nm & Yes & No & 1-2.7 & 2 & 0-80 & No & No & No   \\
		\hline
		
		Qian, et al. \cite{qian2017sidido} & 250nm & No & Yes & 0.5-2 & 0-5 & 5-1000 & No & No & No  \\ 
		\hline
		
		Shih, et al. \cite{shih2011inductorless} & 130nm & Yes & Yes & 1.8 & 1.4 & $<$ 10 & No & No & No  \\
		\hline
		
		Kim, et al.	 \cite{kim2013energy} & 350nm & No & Yes & 1.5-5 & 0-4 & 800 & No & No & No\\
		\hline
		
		Liu, et al.	\cite{liu2015highly} & 180nm & Yes & Yes & 1-1.5 & 3-3.5 & 0-29 & No & No & No \\
		\hline
		
Ram, et al. \cite{ram2020eternal} (Eternal-Thing) & 90nm & Yes & Yes & 1-1.5 & 3-3.55 & 0-22 & Yes & Yes & No \\	
		\hline
		
\textbf{Current Paper (Eternal-Thing 2.0)} & 90nm & Yes & Yes & 1-1.5 & 3-3.55 & 0-22 & Yes & Yes & \textbf{Yes}\\	
		\hline
	\end{tabular}
\end{table*}

From the results obtained, it is observed that by replacing conventional RO with aging tolerant RO, ripples at the output reduces. Further, the aging tolerant RO also shows a lower ripple when aging is accelerated due to the temperature rise. The Trojan detection mechanism secures the MPPT module from unwanted reset, thereby improving the reliability of the EHS. As depicted in Fig. \ref{fig:MPPTFINAL2}, the output of the DC-DC converter ($V_{cp\_trip}$ = 3.55 V) exhibits reduced ripples (10mV-11mV) using ARO with reduced $V_{solar}$ and is used to charge the super-capacitor.

\subsection{ Power Consumption }

Fig. \ref{fig:Powerconsumption_DIE} represents the power consumed by each modules in the EHS. The RO and current sensor is the primary power-consuming modules among all, which consume 80\% of total power. The continuous oscillating mode of RO is responsible for its higher power consumption. The current sensor is periodically ON during the MPPT process only to reduce the power consumption.

The power consumed by the PV-EHS is 22$\mu$W (which is $<$ 1mW), which justifies that the harvesting system is designed for ultra-low-power IoT requirements. This PV-EHS is compared with the other state of art solar EHS in Table \ref{tab:STATEOFART}. The EHS, designed for a temperature sensor and wireless trans-receiver CC2500, consumes a maximum power within 22$\mu$W. The die photograph for the reliable EHS chip given for fabrication is shown in Fig. \ref{fig:Powerconsumption_DIE1}.


\section{Conclusion and Future Work}
\label{sec:Conclusion}

The failure due to supply of a sensor node could be a ruinous state of affairs. The denial of service sort attack might cause information loss in IoT. The reliable Analog Trojan resilient solar energy harvesting system is a state of art technology outcome towards clean energy and handling of IoT end-node devices. The solar energy harvesting system (PV-EHS) designed is similar temperament for a minimum voltage of 1.22V as MPP within the vary of 1-1.5V. The ensuing output is 3-3.55V, that is the demand of the many IoT edge node devices. The efficiency of the DC-DC converter is in the range of 87\%- 97\%. The entire module is powered by the solar cell and the higher bias voltages needed are generated on-chip. The aging tolerant RO makes the long-run IoT node prone to attacks due to analog Trojans and ripples at the output. The entire solar EHS with the anti-aging concept is self-sustainable, and no external power supplies are needed. The aging tolerant reliable energy harvesting system proposed is consuming power in the ultra-low power realm range, i.e. 22$\mu$W.

The future directions of this analysis embody the following: The aging analysis of other modules of the EHS can be studied further. The sensors used in IoT has different supply requirement, and by using adequate power management techniques, various power supplies can be generated further. We intent to explore adding custom hardware accelerators to our Eternal-Thing to perform blockchain functionalities faster with minimal energy consumption \cite{8977825}. We also intend to explore biosensors in the Eternal-Thing which can be deployed to reduce spread of the pandemic \cite{9085930}.


\section*{Acknowledgments}
This research work is supported by Special Manpower Development Program for Chips to System Design (SMDP-C2SD) of Government of India.

\bibliographystyle{IEEEtran}
\bibliography{Bibliography_Eternal-Thing-2-Analog-Trojan}

\begin{thebibliography}{10}
\providecommand{\url}[1]{#1}
\csname url@samestyle\endcsname
\providecommand{\newblock}{\relax}
\providecommand{\bibinfo}[2]{#2}
\providecommand{\BIBentrySTDinterwordspacing}{\spaceskip=0pt\relax}
\providecommand{\BIBentryALTinterwordstretchfactor}{4}
\providecommand{\BIBentryALTinterwordspacing}{\spaceskip=\fontdimen2\font plus
\BIBentryALTinterwordstretchfactor\fontdimen3\font minus
  \fontdimen4\font\relax}
\providecommand{\BIBforeignlanguage}[2]{{%
\expandafter\ifx\csname l@#1\endcsname\relax
\typeout{** WARNING: IEEEtran.bst: No hyphenation pattern has been}%
\typeout{** loaded for the language `#1'. Using the pattern for}%
\typeout{** the default language instead.}%
\else
\language=\csname l@#1\endcsname
\fi
#2}}
\providecommand{\BIBdecl}{\relax}
\BIBdecl

\bibitem{omairi2017power}
A.~Omairi, Z.~H. Ismail, K.~A. Danapalasingam, and M.~Ibrahim, ``{Power
  harvesting in wireless sensor networks and its adaptation with maximum power
  point tracking: Current technology and future directions},'' \emph{IEEE
  Internet of Things Journal}, vol.~4, no.~6, pp. 2104--2115, 2017.

\bibitem{ram2020energy}
S.~K. Ram, B.~B. Das, K.~Mahapatra, S.~P. Mohanty, and U.~Choppali, ``{Energy
  Perspectives in IoT Driven Smart Villages and Smart Cities},'' \emph{IEEE
  Consumer Electronics Magazine}, vol.~10, no.~3, May 2021.

\bibitem{shao2009design}
H.~Shao, C.-Y. Tsui, and W.-H. Ki, ``{The design of a micro power management
  system for applications using photovoltaic cells with the maximum output
  power control},'' \emph{IEEE Transactions on Very Large Scale Integration
  (VLSI) Systems}, vol.~17, no.~8, pp. 1138--1142, 2009.

\bibitem{ram2019sehs}
S.~K. Ram, B.~B. Das, B.~Pati, C.~R. Panigrahi, and K.~K. Mahapatra, ``{SEHS:
  Solar Energy Harvesting System for IoT Edge Node Devices},'' in
  \emph{Progress in Advanced Computing and Intelligent Engineering}, 2021, pp.
  432--443.

\bibitem{carreon2014boost}
S.~Carreon-Bautista, A.~Eladawy, A.~N. Mohieldin, and E.~S{\'a}nchez-Sinencio,
  ``{Boost converter with dynamic input impedance matching for energy
  harvesting with multi-array thermoelectric generators},'' \emph{IEEE
  Transactions on Industrial Electronics}, vol.~61, no.~10, pp. 5345--5353,
  2014.

\bibitem{shih2011inductorless}
Y.-C. Shih and B.~P. Otis, ``{An Inductorless {DC-DC} Converter for Energy
  Harvesting With a 1.2$ \mu $W$ $ Bandgap-Referenced Output Controller},''
  \emph{IEEE Transactions on Circuits and Systems II: Express Briefs}, vol.~58,
  no.~12, pp. 832--836, 2011.

\bibitem{das2019spatio}
B.~B. Das, P.~Kumar, D.~Kar, S.~K. Ram, K.~S. Babu, and R.~K. Mohapatra, ``A
  spatio-temporal model for {EEG}-based person identification,''
  \emph{Multimedia Tools and Applications}, vol.~78, no.~19, pp.
  28\,157--28\,177, 2019.

\bibitem{ram2019ultra}
S.~K. Ram, B.~B. Das, A.~K. Swain, and K.~K. Mahapatra, ``Ultra-low power solar
  energy harvester for {IoT} edge node devices,'' in \emph{2019 IEEE
  International Symposium on Smart Electronic Systems (iSES)(Formerly iNiS)},
  2019, pp. 205--208.

\bibitem{kim2011regulated}
J.~Kim, J.~Kim, and C.~Kim, ``{A regulated charge pump with a low-power
  integrated optimum power point tracking algorithm for indoor solar energy
  harvesting},'' \emph{IEEE Transactions on Circuits and Systems II: Express
  Briefs}, vol.~58, no.~12, pp. 802--806, 2011.

\bibitem{liu2015highly}
X.~Liu and E.~S{\'a}nchez-Sinencio, ``{A highly efficient ultralow photovoltaic
  power harvesting system with {MPPT} for internet of things smart nodes},''
  \emph{IEEE Transactions on Very Large Scale Integration (VLSI) systems},
  vol.~23, no.~12, pp. 3065--3075, 2015.

\bibitem{8287053}
S.~S. {Roy}, D.~{Puthal}, S.~{Sharma}, S.~P. {Mohanty}, and A.~Y. {Zomaya},
  ``Building a sustainable internet of things: Energy-efficient routing using
  low-power sensors will meet the need,'' \emph{IEEE Consumer Electronics
  Magazine}, vol.~7, no.~2, pp. 42--49, March 2018.

\bibitem{7539244}
S.~P. {Mohanty}, U.~{Choppali}, and E.~{Kougianos}, ``Everything you wanted to
  know about smart cities: The internet of things is the backbone,'' \emph{IEEE
  Consumer Electronics Magazine}, vol.~5, no.~3, pp. 60--70, July 2016.

\bibitem{zanella2014internet}
A.~Zanella, N.~Bui, A.~Castellani, L.~Vangelista, and M.~Zorzi, ``{Internet of
  things for smart cities},'' \emph{IEEE Internet of Things journal}, vol.~1,
  no.~1, pp. 22--32, 2014.

\bibitem{gubbi2013internet}
J.~Gubbi, R.~Buyya, S.~Marusic, and M.~Palaniswami, ``{Internet of Things
  (IoT): A Vision, Architectural Elements, and Future Directions},''
  \emph{Future Generation Computer Systems}, vol.~29, no.~7, pp. 1645--1660,
  September 2013.

\bibitem{rostirolla2017elcity}
G.~Rostirolla, R.~da~Rosa~Righi, J.~L.~V. Barbosa, and C.~A. da~Costa,
  ``{ElCity: An elastic multilevel energy saving model for smart cities},''
  \emph{IEEE Transactions on Sustainable Computing}, vol.~3, no.~1, pp. 30--43,
  2017.

\bibitem{Alladi_MCE_2020-Mar}
T.~{Alladi}, V.~{Chamola}, B.~{Sikdar}, and K.~R. {Choo}, ``{Consumer {IoT}:
  Security Vulnerability Case Studies and Solutions},'' \emph{IEEE Consumer
  Electronics Magazine}, vol.~9, no.~2, pp. 17--25, March 2020.

\bibitem{9333580}
D.~{Puthal}, S.~P. {Mohanty}, S.~{Wilson}, and U.~{Choppali}, ``Collaborative
  edge computing for smart villages,'' \emph{IEEE Consumer Electronics
  Magazine}, vol.~10, no.~3, May 2021.

\bibitem{davies2020iot}
A.~Davies, ``{IoT, Smart Technologies, Smart Policing: The Impact for Rural
  Communities},'' in \emph{Smart Village Technology}.\hskip 1em plus 0.5em
  minus 0.4em\relax Springer, 2020, pp. 25--37.

\bibitem{9153927}
P.~{Chanak} and I.~{Banerjee}, ``Internet of things-enabled smart villages:
  Recent advances and challenges,'' \emph{IEEE Consumer Electronics Magazine},
  vol.~10, no.~3, May 2021.

\bibitem{8977808}
S.~P. {Mohanty}, ``Security and privacy by design is key in the internet of
  everything ({IoE}) era,'' \emph{IEEE Consumer Electronics Magazine}, vol.~9,
  no.~2, pp. 4--5, March 2020.

\bibitem{1512191}
S.~P. {Mohanty}, N.~{Ranganathan}, and R.~K. {Namballa}, ``A {VLSI}
  architecture for visible watermarking in a secure still digital camera
  (s$^2$dc) design,'' \emph{IEEE Transactions on Very Large Scale Integration
  (VLSI) Systems}, vol.~13, no.~8, pp. 1002--1012, August 2005.

\bibitem{MOHANTY2009468}
S.~P. Mohanty, ``A secure digital camera architecture for integrated real-time
  digital rights management,'' \emph{Journal of Systems Architecture}, vol.~55,
  no.~10, pp. 468--480, October-December 2009.

\bibitem{8263198}
S.~P. {Mohanty}, E.~{Kougianos}, and P.~{Guturu}, ``{SBPG}: Secure better
  portable graphics for trustworthy media communications in the {IoT},''
  \emph{IEEE Access}, vol.~6, pp. 5939--5953, 2018.

\bibitem{1632351}
S.~P. {Mohanty}, N.~{Ranganathan}, and K.~{Balakrishnan}, ``A dual
  voltage-frequency {VLSI} chip for image watermarking in {DCT} domain,''
  \emph{IEEE Transactions on Circuits and Systems II: Express Briefs}, vol.~53,
  no.~5, pp. 394--398, May 2006.

\bibitem{8752409}
V.~P. {Yanambaka}, S.~P. {Mohanty}, E.~{Kougianos}, and D.~{Puthal}, ``{PMsec}:
  Physical unclonable function-based robust and lightweight authentication in
  the internet of medical things,'' \emph{IEEE Transactions on Consumer
  Electronics}, vol.~65, no.~3, pp. 388--397, August 2019.

\bibitem{ram2020eternal}
S.~K. {Ram}, S.~R. {Sahoo}, B.~B. {Das}, K.~{Mahapatra}, and S.~P. {Mohanty},
  ``{Eternal-Thing}: A secure aging-aware solar-energy harvester thing for
  sustainable {IoT},'' \emph{IEEE Transactions on Sustainable Computing}, no.
  10.1109/TSUSC.2020.2987616, 2020.

\bibitem{guo2019capacitors}
X.~Guo, H.~Zhu, Y.~Jin, and X.~Zhang, ``{When Capacitors Attack: Formal Method
  Driven Design and Detection of Charge-Domain Trojans},'' in \emph{2019
  Design, Automation \& Test in Europe Conference \& Exhibition (DATE)}, 2019,
  pp. 1727--1732.

\bibitem{yang2016a2}
K.~Yang, M.~Hicks, Q.~Dong, T.~Austin, and D.~Sylvester, ``{A2: Analog
  malicious hardware},'' in \emph{2016 IEEE Symposium on Security and Privacy
  (SP)}, 2016, pp. 18--37.

\bibitem{kim2013energy}
H.~Kim, S.~Kim, C.-K. Kwon, Y.-J. Min, C.~Kim, and S.-W. Kim, ``{An
  energy-efficient fast maximum power point tracking circuit in an 800-$\mu$W
  photovoltaic energy harvester},'' \emph{IEEE Transactions on Power
  Electronics}, vol.~28, no.~6, pp. 2927--2935, 2013.

\bibitem{mondal2017chip}
S.~Mondal and R.~Paily, ``On-chip photovoltaic power harvesting system with
  low-overhead adaptive {MPPT} for {IoT} nodes,'' \emph{IEEE Internet of Things
  Journal}, vol.~4, no.~5, pp. 1624--1633, 2017.

\bibitem{ram2018energy}
S.~K. Ram, S.~R. Sahoo, K.~Sudeendra, and K.~Mahapatra, ``Energy efficient
  ultra low power solar harvesting system design with {MPPT} for {IoT} edge
  node devices,'' in \emph{IEEE International Symposium on Smart Electronic
  Systems (iSES)(Formerly iNiS)}, 2018, pp. 130--133.

\bibitem{ram2020solar}
S.~K. {Ram}, S.~{Chourasia}, B.~B. {Das}, A.~K. {Swain}, K.~{Mahapatra}, and
  S.~P. {Mohanty}, ``{A Solar Based Power Module for Battery-Less IoT Sensors
  Towards Sustainable Smart Cities},'' in \emph{2020 IEEE Computer Society
  Annual Symposium on VLSI (ISVLSI)}, 2020, pp. 458--463.

\bibitem{estrada2019fully}
J.~J. Estrada-L{\'o}pez, A.~Abuellil, A.~Costilla-Reyes, M.~Abouzied, S.~Yoon,
  and E.~S{\'a}nchez-Sinencio, ``{A fully integrated maximum power tracking
  combiner for energy harvesting IoT applications},'' \emph{IEEE Transactions
  on Industrial Electronics}, vol.~67, no.~4, pp. 2744--2754, 2019.

\bibitem{abuellil2019multiple}
A.~Abuellil, J.~J. Estrada-L{\'o}pez, A.~Bommireddipalli, A.~Costilla-Reyes,
  Z.~Zeng, and E.~S{\'a}nchez-Sinencio, ``{Multiple-Input Harvesting Power
  Management Unit With Enhanced Boosting Scheme for IoT Applications},''
  \emph{IEEE Transactions on Industrial Electronics}, vol.~67, no.~5, pp.
  3662--3672, 2019.

\bibitem{ma2019sensing}
D.~Ma, G.~Lan, M.~Hassan, W.~Hu, and S.~K. Das, ``{Sensing, computing, and
  communications for energy harvesting IoTs: A survey},'' \emph{IEEE
  Communications Surveys \& Tutorials}, vol.~22, no.~2, pp. 1222--1250, 2019.

\bibitem{chew2018power}
Z.~J. Chew, T.~Ruan, and M.~Zhu, ``{Power management circuit for wireless
  sensor nodes powered by energy harvesting: On the synergy of harvester and
  load},'' \emph{IEEE Transactions on Power Electronics}, vol.~34, no.~9, pp.
  8671--8681, 2018.

\bibitem{chang2018modeling}
Y.-H. Chang and J.-S. Lin, ``{Modeling and implementation of high-gain
  coupled-inductor switched-capacitor step-up DC-DC converter},'' in \emph{2018
  3rd International Conference on Control and Robotics Engineering
  (ICCRE)}.\hskip 1em plus 0.5em minus 0.4em\relax IEEE, 2018, pp. 135--138.

\bibitem{sengupta2018supercapacitors}
A.~S. Sengupta, S.~Satpathy, S.~P. Mohanty, D.~Baral, and B.~K. Bhattacharyya,
  ``{Supercapacitors Outperform Conventional Batteries [Energy and
  Security]},'' \emph{IEEE Consumer Electronics Magazine}, vol.~7, no.~5, pp.
  50--53, 2018.

\bibitem{biziitu2017improving}
F.~B{\^\i}z{\^\i}itu and L.~Gora{\c{s}}, ``{Improving IC power efficiency by
  implementing charge recycling in Dickson charge pumps with multiple pumping
  branches},'' in \emph{2017 International Semiconductor Conference
  (CAS)}.\hskip 1em plus 0.5em minus 0.4em\relax IEEE, 2017, pp. 187--190.

\bibitem{ballo2019review}
A.~Ballo, A.~D. Grasso, and G.~Palumbo, ``{A review of charge pump topologies
  for the power management of IoT nodes},'' \emph{Electronics}, vol.~8, no.~5,
  p. 480, 2019.

\bibitem{qian2017sidido}
Y.~Qian, H.~Zhang, Y.~Chen, Y.~Qin, D.~Lu, and Z.~Hong, ``A {SIDIDO} {DC-DC}
  converter with dual-mode and programmable-capacitor-array {MPPT} control for
  thermoelectric energy harvesting,'' \emph{IEEE Transactions on Circuits and
  Systems II: Express Briefs}, vol.~64, no.~8, pp. 952--956, 2017.

\bibitem{yue2019charge}
X.~Yue, J.~Kiely, D.~Gibson, and E.~M. Drakakis, ``{Charge-based supercapacitor
  storage estimation for indoor sub-mW photovoltaic energy harvesting powered
  wireless sensor nodes},'' \emph{IEEE Transactions on Industrial Electronics},
  vol.~67, no.~3, pp. 2411--2421, 2019.

\bibitem{zhang2013design}
X.~Zhang and M.~Tehranipoor, ``{Design of on-chip lightweight sensors for
  effective detection of recycled ICs},'' \emph{IEEE transactions on very large
  scale integration (VLSI) systems}, vol.~22, no.~5, pp. 1016--1029, 2013.

\bibitem{guin2015design}
U.~Guin, D.~Forte, and M.~Tehranipoor, ``{Design of accurate low-cost on-chip
  structures for protecting integrated circuits against recycling},''
  \emph{IEEE Transactions on Very Large Scale Integration (VLSI) Systems},
  vol.~24, no.~4, pp. 1233--1246, 2015.

\bibitem{sahoo2018novel}
S.~R. Sahoo, S.~Kumar, and K.~Mahapatra, ``A novel configurable ring oscillator
  {PUF} with improved reliability using reduced supply voltage,''
  \emph{Microprocessors and Microsystems}, vol.~60, pp. 40--52, 2018.

\bibitem{jaw2012analysis}
B.-Y. Jaw and H.~Lin, ``An analysis of output ripples for {PMOS} charge pumps
  and design methodology,'' in \emph{2012 IEEE Asia Pacific Conference on
  Circuits and Systems}, 2012, pp. 424--427.

\bibitem{rahman2016aging}
M.~T. Rahman, F.~Rahman, D.~Forte, and M.~Tehranipoor, ``An aging-resistant
  {RO-PUF} for reliable key generation,'' \emph{IEEE Transactions on Emerging
  Topics in Computing}, vol.~4, no.~3, pp. 335--348, 2016.

\bibitem{8977825}
S.~P. {Mohanty}, V.~P. {Yanambaka}, E.~{Kougianos}, and D.~{Puthal},
  ``{PUFchain}: A hardware-assisted blockchain for sustainable simultaneous
  device and data security in the internet of everything ({IoE}),'' \emph{IEEE
  Consumer Electronics Magazine}, vol.~9, no.~2, pp. 8--16, May 2020.

\bibitem{9085930}
A.~K. {Tripathy}, A.~G. {Mohapatra}, S.~P. {Mohanty}, E.~{Kougianos}, A.~M.
  {Joshi}, and G.~{Das}, ``{EasyBand}: A wearable for safety-aware mobility
  during pandemic outbreak,'' \emph{IEEE Consumer Electronics Magazine},
  vol.~9, no.~5, pp. 57--61, Sep 2020.

\end{thebibliography}

\section*{Authors' Biographies}


\begin{minipage}{\columnwidth}
	\begin{wrapfigure}{l}{1.2in}
		\vspace{-0.3cm}
		\includegraphics[height=1.2in,keepaspectratio]{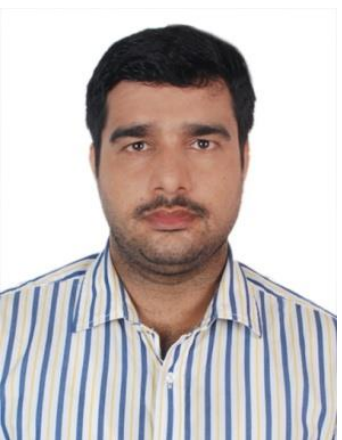}
		\vspace{-0.5cm}
	\end{wrapfigure}
	\noindent
\textbf{Saswat Kumar Ram} (Student Member, IEEE) received the professional degree in Electronics and Telecommunication Engineering from Biju Patanaik University of Technology, India in 2005, the M.Tech degree in VLSI Design and Embedded systems from NIT, Rourkela, India in 2011. He is currently Pursuing the Ph.D. degree in Electronics and communication engineering at NIT, Rourkela, India. He is a Student Member, IEEE and IEEE circuits and systems society. His current research interest includes energy harvesting for IoT, low power VLSI design, embedded systems, Hardware security, signal processing, machine learning.
\end{minipage}

\vspace{0.9cm}

\begin{minipage}{\columnwidth}
	\begin{wrapfigure}{l}{1.2in}
	\vspace{-0.3cm}
	\includegraphics[height=1.2in,keepaspectratio]{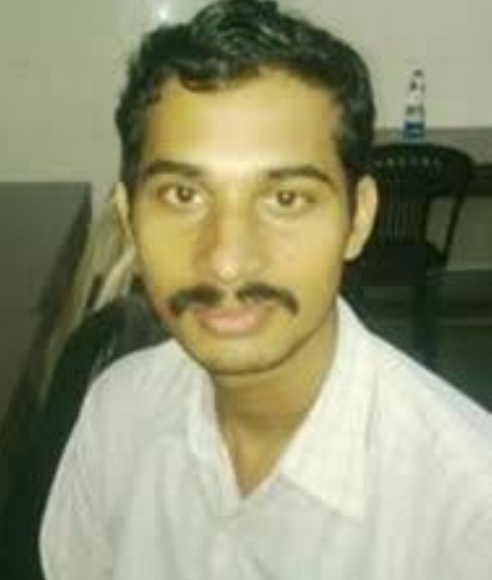}
	\vspace{-0.5cm}
\end{wrapfigure}
\noindent
\textbf{Sauvagya Ranjan Sahoo} (Student Member, IEEE) received the engineering degree (EN\&TC) from DRIEMS, Odisha in 2006. In 2012, he obtained his M.Tech in VLSI Design \& Embedded Systems from National Institute of   Technology, Rourkela, Odisha. In 2019, he obtained his Ph.D.  in Electronics and Communication Engineering from National Institute of Technology, Rourkela. His main research interests include low power CMOS VLSI circuit design, Hardware security.
\end{minipage}

\vspace{0.9cm}

\begin{minipage}{\columnwidth}
	\begin{wrapfigure}{l}{1.2in}
	\vspace{-0.3cm}
	\includegraphics[height=1.2in,keepaspectratio]{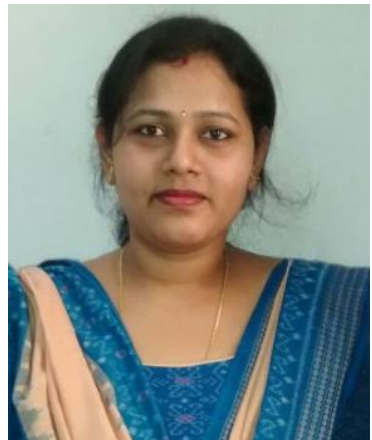}
	\vspace{-0.5cm}
\end{wrapfigure}
\noindent
\textbf{Banee Bandana Das} (Student Member, IEEE) received the professional degree in Computer Science Engineering from Biju Patanaik University of Technology, India in 2010, the M.Tech degree from from Biju Patanaik University of Technology, India in 2012. She is currently Pursuing the Ph.D. degree in computer science engineering in NIT, Rourkela, India. Her current research interest includes machine learning, computer vision, neural network and computational intelligence, Internet-of-Things.
\end{minipage}

\vspace{0.9cm}

\begin{minipage}{\columnwidth}
		\begin{wrapfigure}{l}{1.2in}
		\vspace{-0.3cm}
\includegraphics[height=1.2in,keepaspectratio]{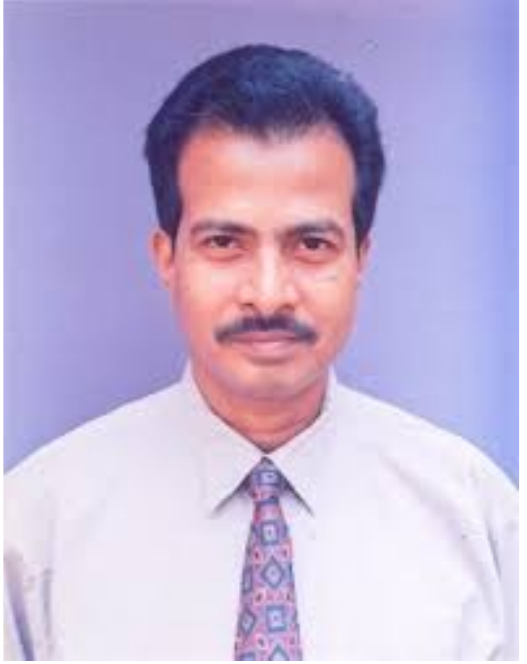}
		\vspace{-0.5cm}
\end{wrapfigure}
\noindent
\textbf{Kamalakanta Mahapatra} (Senior Member, IEEE) obtained his B. Tech degree with Honours from Regional Engineering College, Calicut in 1985, M. Tech from Regional Engineering College, Rourkela in 1989 and Ph. D. from IIT Kanpur in 2000. He is currently a Professor in Electronics and Communication Engineering Department of National Institute of Technology (NIT), Rourkela. He assumed this position since February 2004. He is a fellow of the Institution of Engineers (India) in ECE Division. He has published several research papers in National and International Journals. His research interests include Embedded Computing Systems, VLSI Design, Hardware Security and Industrial Electronics.
\end{minipage}

\vspace{0.9cm}

\begin{minipage}[htbp]{\columnwidth}
	\begin{wrapfigure}{l}{1.2in}
		\vspace{-0.3cm}
		\includegraphics[width=1.2in,keepaspectratio]{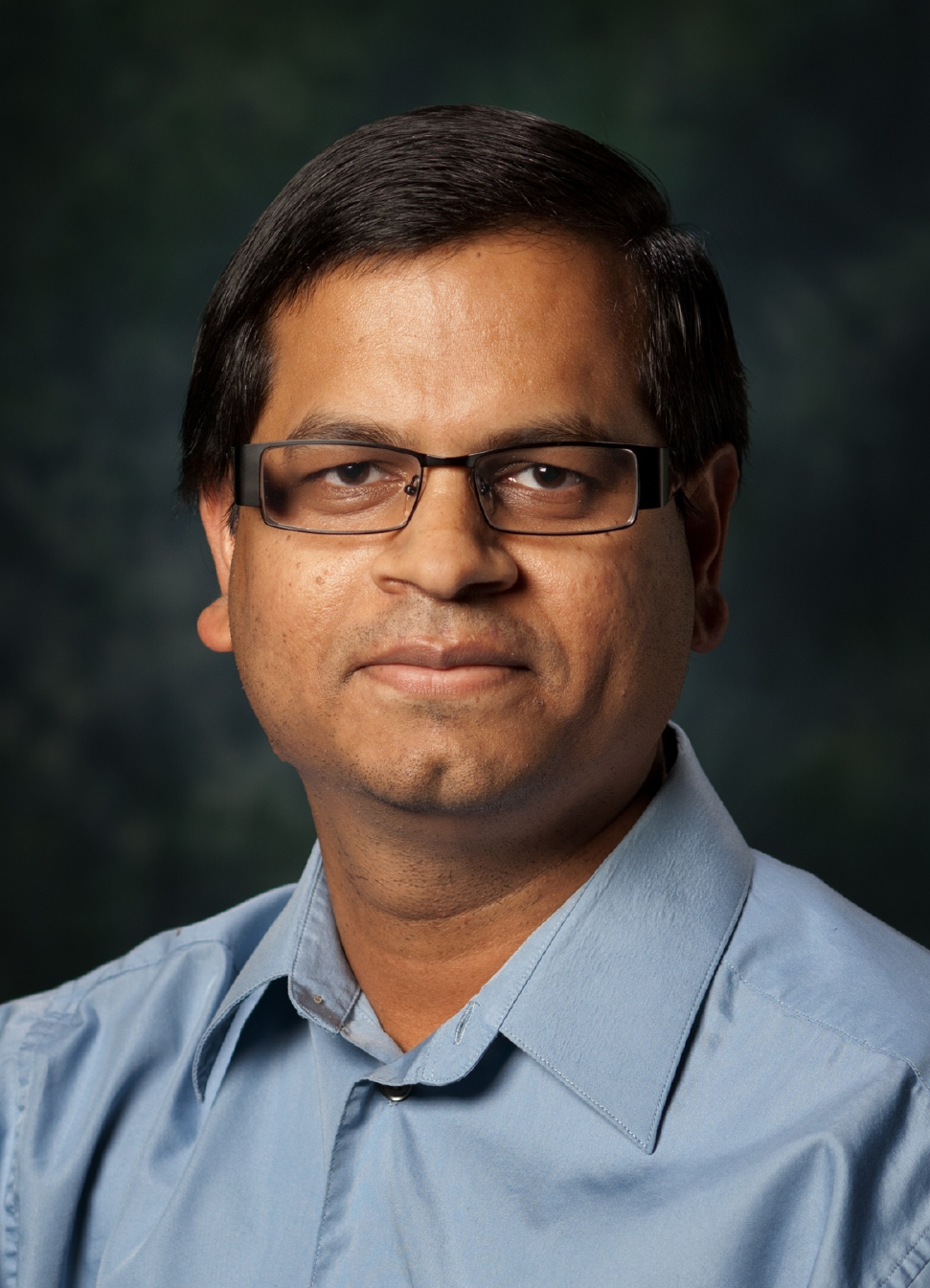}
		\vspace{-0.5cm}
	\end{wrapfigure}
	\noindent
\textbf{Saraju P. Mohanty} (Senior Member, IEEE)  received the bachelor's degree (Honors) in electrical  engineering from Orissa University of Agriculture  and Technology, Bhubaneswar, India, in 1995, the  master's degree in systems science and automation from Indian Institute of Science at Bengaluru, Bengaluru, India, in 1999, and the Ph.D. degree  in computer science and engineering from the University of South Florida, Tampa, FL, USA, in  2003. 

He is a Professor with the University of North Texas, Denton, TX, USA. His research is in ``Smart Electronic Systems'' which has been funded by National Science Foundations, Semiconductor Research Corporation, U.S. Air Force, IUSSTF, and Mission Innovation. He has authored 350 research articles, four books, and invented four granted and three pending patents. His Google Scholar h-index is 40 and i10-index is 149 with 6900 citations. He is regarded as a visionary researcher on smart cities technology in which his research deals with security and energy aware and AI/ML-integrated smart components. He introduced the Secure Digital Camera in 2004 with built-in security features designed using Hardware-Assisted Security or Security by Design principle. He is widely credited as the designer of the first digital watermarking chip in 2004 and first the low-power digital watermarking chip in 2006. He has mentored 2 Postdoctoral Researchers and supervised 13 Ph.D. dissertations, 26 M.S. thesis, and 11 undergraduate projects. 

Prof. Mohanty is a recipient of 13 Best Paper Awards, the Fulbright Specialist Award in 2020, the IEEE Consumer Technology Society Outstanding Service Award in 2020, the IEEE-CS-TCVLSI Distinguished Leadership Award in 2018, and the PROSE Award for Best Textbook in Physical Sciences and Mathematics category in 2016. He has delivered ten keynotes and served on 11 panels with various International Conferences. He has been serving on the editorial board of several peer-reviewed international journals, such as IEEE TRANSACTIONS ON CONSUMER ELECTRONICS, and IEEE TRANSACTIONS ON BIG DATA. He is the Editor-in-Chief of the IEEE Consumer Electronics Magazine. He has been serving on the Board of Governors of the IEEE Consumer Technology Society, and has served as the Chair of Technical Committee on Very Large Scale Integration, IEEE Computer Society (IEEE-CS) from 2014 to 2018. He is the Founding Steering Committee Chair of the IEEE International Symposium on Smart Electronic Systems, the Steering Committee Vice-Chair of the IEEE-CS Symposium on VLSI, and the Steering Committee Vice-Chair of the OITS International  Conference on Information Technology.

\end{minipage}

\end{document}